\documentclass[10pt, journal]{IEEEtran}
\usepackage{amsfonts}
\usepackage{amssymb}
\usepackage[ruled,vlined]{algorithm2e}
\usepackage{mathtools}  
\usepackage{amsthm}
\usepackage[]{algorithm2e}
\usepackage{subfigure}
\usepackage{tabulary}
\usepackage{footnote}
\usepackage[noend]{algpseudocode}
\usepackage{algorithmicx}
\usepackage[export]{adjustbox}
\usepackage{booktabs}
\usepackage[justification=centering]{caption}
\usepackage{comment}
\newtheorem{remark}{Remark}
\usepackage{cases}
\usepackage{cite}
\usepackage{multicol}
\usepackage{xcolor}
\usepackage{graphicx}
\graphicspath{{./}{figures/}}

\usepackage{stfloats}
\usepackage{cuted}
\usepackage{lipsum}  
\DeclareMathOperator*{\argmax}{argmax}

\usepackage{subfigure}

\newtheorem{Prob}{\textbf{Problem}}
\newtheorem{lemma}{Lemma}

\title{ IEEE 802.11be Network Throughput Optimization with Multi-Link Operation and AP Controller} 

 \author{\text{Lyutianyang Zhang}, \text{Hao Yin},  \text{Sumit Roy}, \textit{Fellow,~IEEE}, \text{Liu Cao}, \text{Xiangyu Gao}, \text{Vanlin Sathya}
 \thanks{Lyutianyang Zhang, Hao Yin, Sumit Roy, Liu Cao, and Xiangyu Gao are with Department of Electrical \& Computer Engineering, University of Washington, Seattle, WA, USA (e-mail:\{lyutiz, haoyin, sroy, liucao, xygao\}@uw.edu. V. Sathya is with the Department of Computer
 Science, University of Chicago, Chicago, IL 60637 USA (e-mail:
 vanlin@uchicago.edu). (\emph{Corresponding author: Liu Cao})}}
\begin{document}

\maketitle

\begin{abstract}

IEEE 802.11be (Wi-Fi 7) introduces a new concept called multi-link operation (MLO), which allows multiple Wi-Fi interfaces in different bands (2.4, 5, and 6 GHz) to work together to increase network throughput, reduce latency, and improve spectrum reuse efficiency in dense overlapping networks. To make the most of MLO, this paper proposes a new data-driven resource allocation algorithm for the 11be network with the aid of an access point (AP) controller. To maximize network throughput, a network topology optimization problem is formulated for 11be network, which is solved by exploiting the totally unimodular property of the bipartite graph formed by the connection between AP and station (STA) in Wi-Fi networks. Subsequently, a proportional fairness algorithm is applied for radio link allocation, network throughput optimization considering the channel condition, and the fairness of the multi-link device (MLD) data rate. The performance of the proposed algorithm on two main MLO implementations - multi-link multi-radio (MLMR) with simultaneous transmission and reception (STR), and the interplay between multiple nodes employing them are evaluated through cross-layer (PHY-MAC) data rate simulation with PHY abstraction.
\end{abstract}


\section{Introduction} 
\label{introduction}

\IEEEPARstart{A}{s} tracked by the Cisco Annual Internet Report \cite{Cisco}, there will be an average of 3.6 devices per capita by 2023 and 5.3 billion global internet users. Among these, exponentiating demand for multi-media streaming, e.g., 4K/8K video \cite{Perez2019AP}, online gaming, and other cloud-enabled services, will require network capacity to extend beyond current generation (Wi-Fi 6) limits of $10$ Gbps peak. Additionally, future Wi-Fi network standards must also satisfy user experience requirements with respect to link reliability, worst-case latency, and extremely high speed. Currently, Wi-Fi 6/6E WLANs based on the IEEE 802.11ax amendment \cite{Cisco} are already commercially available, and new standardization efforts are now concentrating on the IEEE 802.11be (we refer it to as 11be for simplicity) amendment \cite{Perez2019AP} for future Wi-Fi 7. The allocation of the entire 6 GHz band (5.925-7.125 GHz) in the US, South Korea, etc., and a lower 480 MHz band in Europe (5945-6425 MHz) for unlicensed enable freedom to explore novel greenfield system architectures and protocol stack concepts, as illustrated in this paper.


IEEE 802.11be extremely high throughput (EHT) \cite{Perez2019AP} is
targeted for future dense use cases - mainly in enterprises and
other high-density (hotspot-like) environments, expected to
serve the increasing demand for high-definition multimedia
traffic. Furthermore, emerging applications such as artificial/virtual reality (AR/VR), real-time gaming, and edge cloud
computing services are well known to require both
extremely high throughput and high
link reliability \cite{Perez2019AP}. Hence, beyond increasing PHY data rates
via a combination of higher modulation orders, large channel
bandwidths, and an increasing number of spatial streams, new MAC
features such as multi-link operation (MLO) and multi-access
point coordination (MAPC) \cite{9770579, zhangwifi} have been proposed.

R2: MLO promotes the simultaneous use of multiple wireless interfaces for concurrent data transmission and reception at access points (APs) and stations (STAs) via dual- or tri-band radio capabilities. In hotspots or dense cells, interference-induced channel loss and/or overlapping transmissions-induced collisions occur more frequently, especially when each AP (or cell) has no prior knowledge of other APs (or cells) on the current network status. Thus, it is desirable to enable some degree of collaboration among neighboring APs through new control frames for inter-AP information exchange among coordination set \cite{zhangwifi, ahn2020novel}. Although MLO has been well known for providing higher reliability and latency in next-generation Wi-Fi \cite{8928021,9557495}, MLO itself is not enough without AP coordination; therefore, a coordinated AP architecture that enables the full potential (EHT, reduced worst-case latency, and increased link stability) of MLO is necessary. 

%

\begin{figure}
    \centering
    \includegraphics[height=0.25\textwidth, width=.36\textwidth]{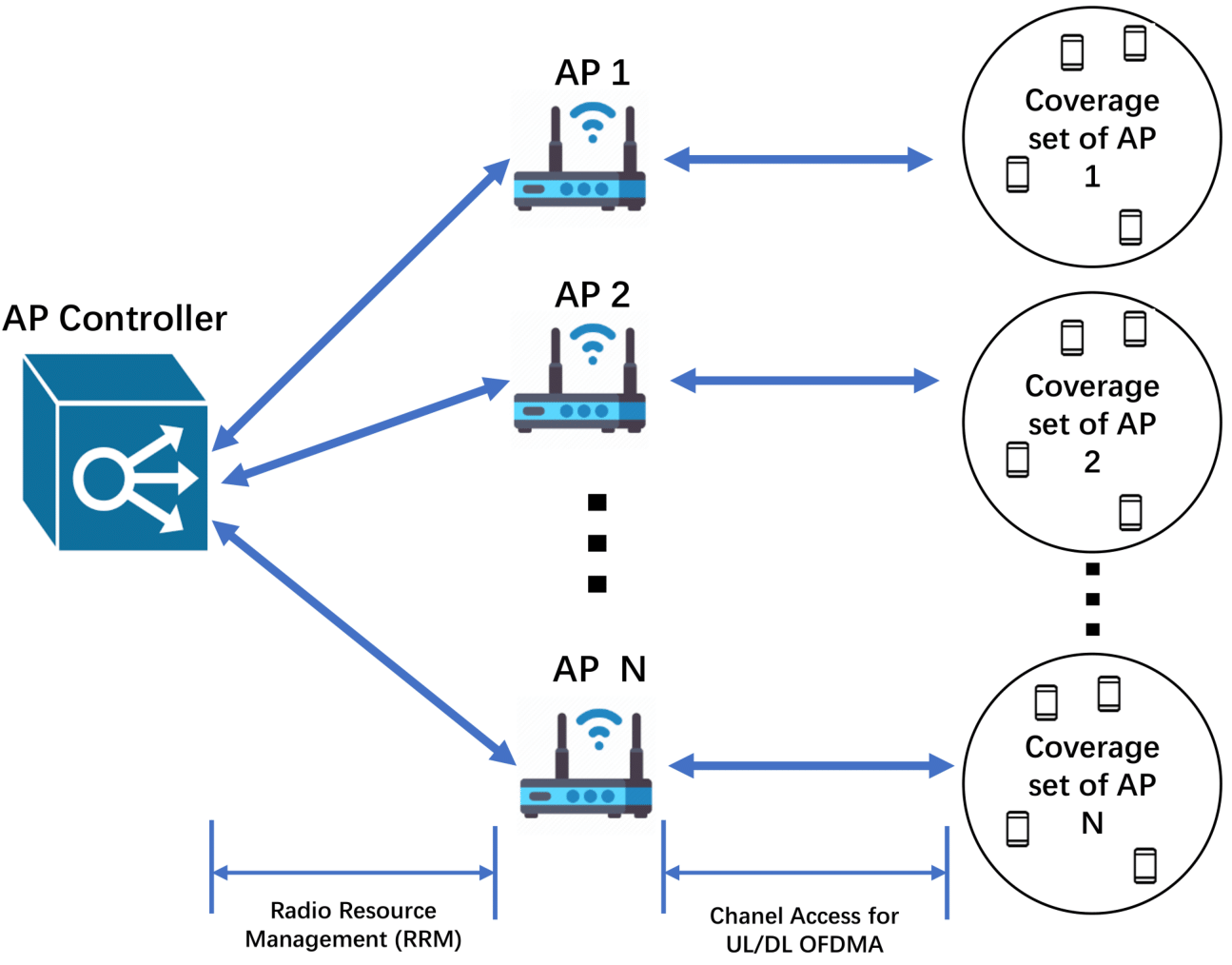}
    \caption{Proposed Architecture for MAPC.}
    \label{fig:APC}
    \vspace{-0.5cm}
\end{figure}

The system architecture in Fig.\ref{fig:APC} depicts the enterprise multi-AP coordination, whereby all APs share information with the controller regarding resource allocation, data rate, channel state information (CSI), etc. Resource allocation is expected to produce lower collision probability, higher throughput, and better worst-case latency for channel access. The centralized AP controller (APC) \footnote{R1\_1: Only one APC is introduced in our proposed architecture because enterprise AP controller can manage up to hundreds of AP and the corresponding data should all be centralized on APC as the input to our proposed algorithm.} implements channel configuration, i.e., assigning AP-STA pairing and radio link allocation with the consideration of proportional fairness (PF). This is part of the overall management of APC radio resources (RRM) \cite{7727996}, which R2: includes monitoring traffic, achieving throughput and latency of user access, and periodically reconfiguring settings to improve network efficiency. Each multi-link device (MLD) is able to contend for uplink (UL)/downlink (DL) transmission opportunity (TXOP) on its assigned channels utilized for communications with the aid of the novel MAPC architecture proposed in the previous art \cite{zhangwifi}. In this paper, we propose an optimized paradigm for the 11be network with MLO and MAPC, considering the network throughput and fairness of the allocation of radio links.


A dynamic strategy such as multi-link congestion-aware load balancing (MCAB) has been proposed to periodically adjust the traffic-to-link allocation in order to follow channel occupancy changes. The traffic distribution policy over multiple links \cite{9770579, 9765523, carrascosa2022performance} has been experimentally demonstrated to improve worst-case latency and network throughput. Although MCAB is proposed to improve any existing MLO topology, these works have not been studied on intelligent AP-STA pairing and radio link allocation problems. Furthermore, global information about overlapping APs and STAs such as historical throughput, channel condition, and the number of STAs for each AP can be utilized to support the 11be network optimization. In our paper, we propose an AP-STA pairing and radio link allocation for enterprise-level network architecture. It should be noted that the optimization problem is formulated by assuming the coexistence \cite{carrascosa2022performance} of single link (SL), multi-link single radio (MLSR), and multi-link multiple radio (MLMR).

The performance analysis of synchronous MLO is investigated in \cite{9838923}. This work has concluded that the longest backoff window and shortest backoff window should be closer to each other when the number of links grows in the network. In our work, we focus on simultaneous transmission and reception (STR) and optimization of the MLO network topology, considering multiple APs instead of the backoff window design for TXOP contention in synchronous MLO with a single AP. Experiments for MLO latency have also been implemented in \cite{9838765} to show that the symmetric occupancy of all links can significantly reduce network latency, which justifies the importance of the fairness consideration in the MLO radio link allocation problem in our work.

As a closer effort to our work, \cite{7414115} and \cite{8969724} investigate the AP-STA association in Wi-Fi networks without using multi-arm bandit (MAB) and graph theory. The MAB method is not suitable for densely overlapping networks, especially when the channel is dynamic, to which the algorithm is unable to converge. A more practical approach proposed in \cite{8969724} utilizes a graph theory formulation to produce an efficient AP-STA association solution. However, only the single-input single-output (SISO)-based Shannon capacity is used as input to the algorithm. In the 11be network, however, the MLD device is expected to support the multiple-input multiple-output (MIMO) and the input to the algorithm should also consider the contention in each channel, which means that the MAC layer cannot be neglected if we want to formulate a realistic AP-STA association and radio link allocation problem.

\subsection{Contribution}
This paper proposes an algorithm for 11be network that maximizes network throughput while preserving network fairness. The major contributions are listed as below:
\begin{itemize}
\item 
An AP-STA pairing problem based on MAPC \cite{zhangwifi} is formulated. The optimal AP-STA pairing policy of the formulated combinatorial optimization problem can be solved using linear programming with the aid of unimodularity. 
\item We further formulate a radio link allocation problem based on the AP-STA pairing solution considering the fairness among radio links. We also propose an algorithm to solve this problem with guaranteed convergence.
\item Performance evaluation results show that the optimal AP-STA pairing solution plus the proposed radio link allocation algorithm can beat the baseline algorithm up to 32\%.

\end{itemize}

The rest of the paper is organized as follows, 1) we first introduce the system operation details of the 11be network; 2) Then, we formulate the AP-STA pairing model aiming at network throughput maximization, to which an optimal AP-STA pairing solution is designed; 3) The proposed radio link allocation algorithm considering proportional fairness follows the AP-STA pairing solution; 4) Evaluation based on analytical channel data rate model is implemented to analyze the performance of proposed algorithms in terms of data rate, fairness, and convergence. 

\section{MLO System Model and Formulation}
\label{sec:architecture}

In this section, we first introduce the transmission modes of 11be network elements enabled with MLO. Then, we model and solve for the expected network throughput optimization in view of preserving proportional fairness (PF) among network nodes in two steps:
A) We first obtain the optimal AP-STA pairing followed by B) the radio link allocation problem. 
In the end, the system operation is introduced to describe the whole AP-STA pairing and radio link allocation process designed for IEEE 802.11be network throughput, fairness, and latency optimization.

\begin{figure*}[t]
    \centering
    \includegraphics[width = 0.75\textwidth,height=0.20\textwidth]{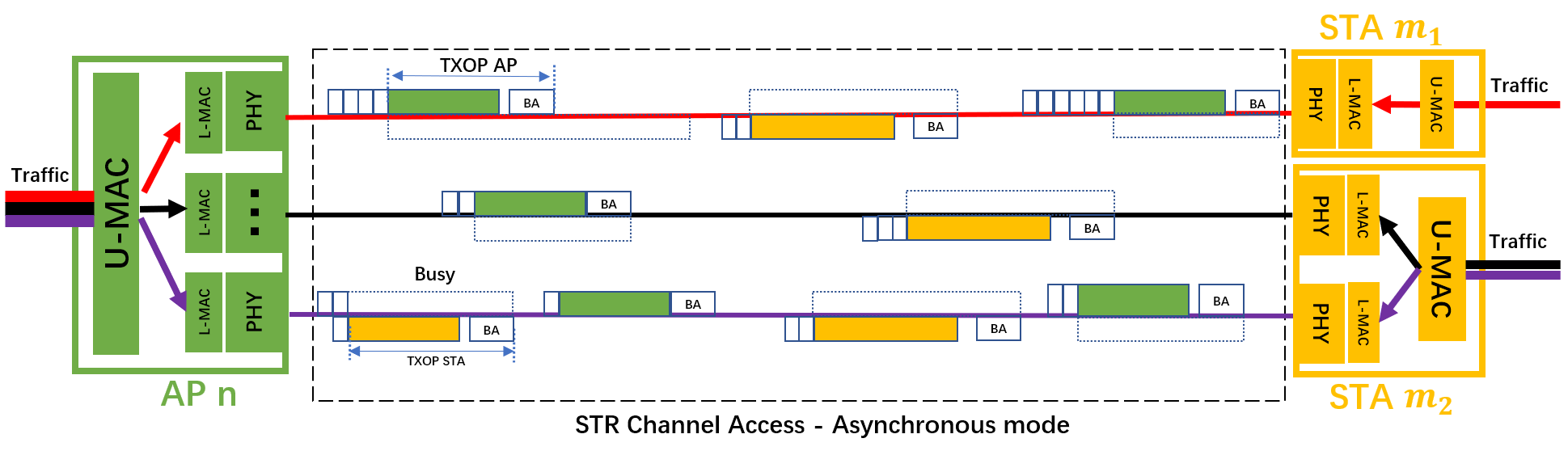}
    \caption{Multi-link Architecture and Transmission Modes.}
    \vspace{-0.5cm}
    \label{fig:MLO}
\end{figure*}
\subsection{Overview of 802.11 MAC and PHY}
The IEEE 802.11 standard~\cite{80211}, a cornerstone of wireless network technology, primarily comprises two layers: the Media Access Control (MAC) layer and the Physical Layer (PHY). The MAC layer serves as the central mediator in wireless communications, responsible for regulating access to the shared communication medium. This layer implements protocols that dictate how devices on a wireless network coordinate their transmissions to avoid collisions and ensure efficient data exchange. Its functions include defining how data packets are framed and addressed, controlling access to the network through mechanisms like carrier sense multiple access with collision avoidance (CSMA/CA) \cite{bianchi2000performance}, and managing error checking to maintain the integrity of the transmitted data. The MAC layer's role is pivotal in maintaining an orderly and efficient communication process, ensuring that multiple devices can share the wireless channel without overwhelming the network.

On the other hand, the PHY layer operates at a more fundamental level, dealing with the actual transmission and reception of raw data bits over the wireless medium. This layer encompasses a range of functions, including the modulation and demodulation of signals, signal transmission and reception, channel frequency selection, and data encryption for security. The PHY layer's specifications vary across different IEEE 802.11 standards (like 802.11a/b/g/n/ac/ax), each tailored to different requirements of transmission speed, coding and modulation scheme, and frequency bands. These variations allow for the adaptation of wireless technology to different environments and uses, from short-range, high-speed data transmission to longer-range communication with more robust signal stability.

\subsection{Multi Link Device (MLD) Enabled Networking}
In the Wi-Fi era preceding 11be such as 11ax, 11ac, etc., each STA is allowed to exchange data with its associated AP in only one of the 2.4, 5, or 6 GHz Wi-Fi bands, i.e., single link operation (SLO). In 802.11be networks with MLO, the aggregation of $2.4$, $5$ and $6$ GHz spectrum allows simultaneous operation between each STA and its associated AP on different bands or channels (orthogonal channel aggregation) with corresponding radio links (PHY interface) \cite{11be_standard}. The MLO architecture with examples of STR and non-STR transmission is shown in Fig.\ref{fig:MLO}. Under STR, each PHY interface maintains its own channel access independent of all others. Such STR capability also enables simultaneous UL and DL communications. 


As shown in Fig. \ref{fig:MLO}, Both AP and STA MLD have MAC and PHY layers. Traffic at AP $n$ goes through a packet split process in U-MAC and is mapped to different radio links corresponding to STA $m_1$ and $m_2$. A traffic identifier (TID), an identifier used to classify a packet in Wireless LANs, can be mapped on UL or DL to a set of enabled links for an MLD. The main function of the Upper MAC (U-MAC) layer of MLD is the selection of the MLD PHY layer for transmission. The main function of Lower MAC (L-MAC) shown in Fig.\ref{fig:MLO} is the channel access for TXOP followed by Block-ACK (BA), control frame and MAC header generation. For each MLD contending for TXOP on any one of its radio links as shown in Fig. \ref{fig:MLO}, it waits until the channel is sensed idle for a distributed inter-frame space (DIFS). Then, a backoff process is initiated. Backoff intervals are slotted, and the discrete backoff time is uniformly distributed in the range [$0, W-1$], where $W$ is the contention window size, and $CW_{min}$ represents the minimum contention window. The backoff counter value is initialized by uniformly choosing an integer from the range [$0, W-1$]. Then, it is decremented by one for each idle slot. The backoff counter is frozen when the channel is sensed busy and will be reactivated until the channel is sensed idle again for a DIFS period. Contention window size $W$ is doubled after each unsuccessful transmission, up to a maximum value $CW_{max}=2^{m}CW_{min}$, where $m$ represents the largest times the contention window size can be doubled.

In this paper, the channel data rate assumed to be known for all radio links between any AP-STA pairs \footnote{In python simulator, the channel data rate is calculated as the total bits of the successfully transmitted packet divided by the time duration in Simulation Section III, which can also be obtained in the same manner in real Wi-Fi system.} is the key input parameter for problem formulation and the optimized data-driven policy for AP-STA pairing and radio link allocation to maximize network throughput with guaranteed proportional fairness in the 11be network. In the system design, we focus on the STR mode, a widely chosen mode over non-STR in industry because of its flexibility and better performance. This assumption allows our simulator design to expand SLO operation to independent transmission and reception on multiple radio links. It is assumed that MLDs are equipped with robust filter design so that in-device coexistence (IDC) does not prevent the 11be network from simultaneous transmission and reception on tri-band, otherwise the transmission on one radio link will interfere the reception on the other radio link.

\begin{figure}[t]
    \centering
    \includegraphics[width = 0.42\textwidth,height=0.24\textwidth]{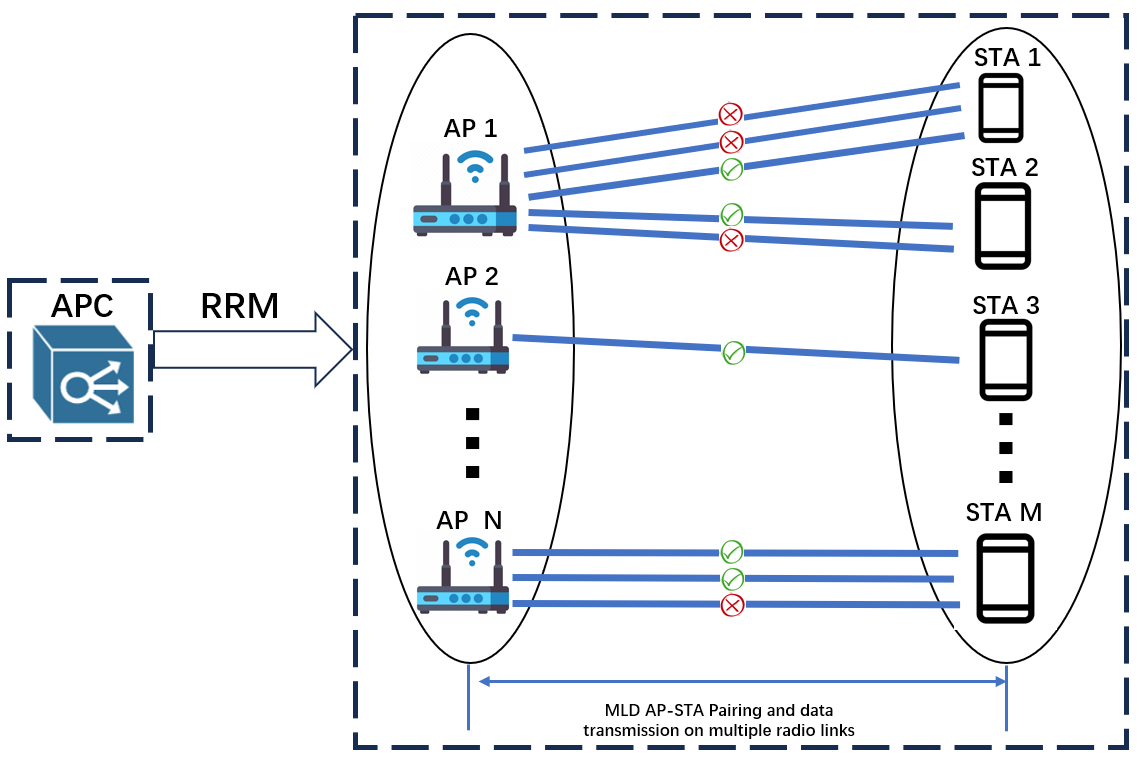}
    \caption{ AP-STA Pairing and Radio Link Allocation.}
    \vspace{-0.5cm}
    \label{fig:ap_sta_graph}
\end{figure}
Suppose that there are $N$ AP MLD and $M$ STA MLD, as shown in Fig.\ref{fig:MLO}. The MAC data plane architecture of AP $n$ has a total of $R(n)$ radio links, and STA $m$ has a total of $r(m)$ radio links. The minimum number of radio links available between any pair of AP and STA is $\min\{R(n),r(m)\}$. An AP MLD or a non-AP MLD is denoted as a vertex, and the connection between any pair of two vertices is denoted as an undirected edge. In the example of radio link allocation shown in Fig. \ref{fig:ap_sta_graph}, $\min\{R(1),r(1)\} = 3$ and only the third radio link is allocated, $\min\{R(1),r(2)\} = 2$ have at most $2$ radio links and only the first radio link is allocated, AP $N$ and STA $M$ have at most 3 radio links and only two radio links are allocated. Each AP MLD can support radio link connection for multiple non-AP MLD while any non-AP MLD can only connect to one AP MLD, where each link has a corresponding maximum data rate corresponding to its channel condition and the performance of IEEE 802.11 distributed coordination function (DCF). If there is no AP-AP or STA-STA data transmission, any topology of the multi-MLD MLO network can be represented by a bipartite graph characterized by an adjacency matrix $\mathbf{A} \in R^{N \times M}$, where the real value $A_{n,m}$ denotes the capacity from vertex $n$ to vertex $m$. In the following, we propose a two-step solution for the optimized multi-MLD MLO network topology, which emphasizes the AP-STA pairing in the first step and radio link allocation in the second step.



\subsection{Step 1: AP-STA Pairing}
To propose the optimized AP-STA pairing policy, several factors should be taken into account: 1) every STA MLD should be connected with an AP MLD; 2) each AP should choose STA MLDs for connection with the highest achievable data rate considering the spectral efficiency and the performance of 802.11 DCF. The channel data rate matrix $\mathbf{C} \in \mathcal{R}^{|\mathcal{F}| \times |\mathbf{E}|}$ denotes the achievable data rate on channel $f \in \mathcal{F}$ for the radio link represented by the edge $e\in \mathbf{E}$ in units of bits/sec/edge. It should be noted that $|\mathbf{E}| = N \times M$; therefore, the channel data rate matrix can also be formulated as $\mathbb{C} \in \mathcal{R}^{|\mathcal{F}| \times N \times M}$. This matrix can only be obtained if each radio link of all available channels is connected with the backlogged throughput information at AP, which is unrealistic. Therefore, all the formulated problems in this paper are solved by iterative algorithms with continuously updated $\mathbf{C}$, the initial value of which can be obtained by taking the standard STA association as input\footnote{The IEEE 802.11 standard utilizes the highest received signal strength indicator (RSSI) of the beacons STA receives to decide to associate with the corresponding AP.}. 

To formulate a problem aiming to maximize aggregate throughput, two variables should be introduced:$\{\mathbb{X}\}$, where $\mathbb{X} \in \{0,1\}^{F \times N \times M}$ and $\mathbf{S} \in \{0,1\}^{F \times |E|}$. $\mathbb{X}_{f.n.m} = 1$ denotes that AP $n$ and STA $m$ are paired on channel $F$ that corresponds to a certain radio link between AP $n$ and STA $m$ by design and $0$ otherwise. Assuming each channel is independent of the other and is chosen for MLO with equal probability i.i.d., the AP-STA pairing problem can be formulated as follows:
\begin{Prob} (AP-STA Pairing + Radio link Allocation: Joint Formulation)
    \begin{align}
     &\argmax_{\mathbb{X}}  \sum_{f=1}^{F}\sum_{n=1}^{N}\sum_{m=1}^{M}\mathbb{X}_{f.n.m}\mathbb{C}_{f,n,m}\\
     s.t.~~&~~\mathbb{X} \in \{0,1\}^{F \times N \times M},\\
         &~~\sum_{f=1}^{F}\sum_{n=1}^{N} \mathbf{X}_{f,n,m} \leq r(m),\forall m \label{eq:sta_mmkp},\\
        &~~\sum_{f=1}^{F} \textbf{card}(\mathbf{X}_{f,n}) \leq 1,\forall m \label{eq:sta_card_mmkp},\\
         &~~\sum_{f=1}^{F}\sum_{m=1}^{N} \mathbf{X}_{f,n,m} \leq R(n),\forall n. \label{eq:ap_mmkp}
    \end{align}
\end{Prob}
The joint formulation for AP-STA Pairing + Radio link Allocation can be generalized as a multiple-choice multi-dimensional knapsack problem (MMKP), which is a well-known NP-hard problem without a solution of polynomial time. Objective (1) represents the maximization of aggregated network throughput. Eq. \eqref{eq:sta_card_mmkp} represents that each STA can be serviced with up to one AP, where $\textbf{card}(\bullet)$ denotes the carnality of vector. Eq. \eqref{eq:sta_mmkp} represents STA $m$ can have a maximum of $r(m)$ operating radio links while Eq. \eqref{eq:ap_mmkp} represents AP $n$ can have a maximum of $R(n)$ operating radio links. As previous arts have discussed \cite{zhang_tcom,hao_tcom}, even a compromised algorithm with a feasible solution is of high time complexity and frequency iterations, which makes it impractical to deploy in 11be networks, refer to Section III.D. Moreover, this formulation cannot achieve fairness among radio links because it aims only at the maximum throughput. Therefore, in this paper, this joint problem is broken into two parts that both have algorithms of low time complexity: 1) AP-STA pairing with variable $\mathbf{X} \in \{0,1\}^{N \times M}$, where $\mathbf{X}_{n,m} = 1$ represents that AP MLD $n$ is paired with STA MLD $m$ and $0$ otherwise (Section II.B); 2) radio link allocation considering proportional fairness with variable $\mathbf{S} \in \{0,1\}^{F \times |E|}$, where $\mathbf{S}_{f,e} = 1$ represents that channel $f$ is chosen as the radio link for the AP-STA pair $e$ and $0$ otherwise (Section II.C). Next, we formulate the first AP-STA pairing problem as follows:
\begin{Prob} (AP-STA Pairing)\label{Prob:AP-STA}
    \begin{align}
     &\argmax_{\mathbf{X}}  \textbf{Tr}(\mathbf{X}^{T}\mathbf{E}_{\mathbf{F}}[\mathbb{C}]) \label{eq:ap_sta_objective}\\
     s.t.~~&~~\mathbf{X} \in \{0,1\}^{N \times M},\\
         &~~\mathbf{1}_{n}^{T}\mathbf{X}_{m} = 1, \forall m \label{eq:sta1}\\
         &~~\mathbf{1}_{m}^{T} \tilde{\mathbf{X}}^{T}_{n} \leq R(n), \forall n \label{eq:apn}.
    \end{align}
\end{Prob}
Objective \eqref{eq:ap_sta_objective} aims to maximize the network throughput considering the average channel data rate for any AP-STA pair. Eq. \eqref{eq:sta1} represents the constraint that each STA is served (at most) by 1 AP and Eq. \eqref{eq:apn} represents that any AP $n$ can support up to $R(n)$ STAs.  $\tilde{\mathbf{X}}^{T}_{n}$ denotes the transpose of the $n_{th}$ row vector of $\mathbf{X}$ and $\mathbf{E}_{\mathcal{F}}[\bullet]$ denotes the expectation over the radio links/channels. The average data rate $D_{n,m}$ across all radio links between AP MLD $n$ and STA MLD $m$ can be expressed as follows:
\begin{equation}
    \begin{aligned}
        \mathbf{D}_{n,m} &= \mathbf{E}_{\mathbf{F}}[\mathbb{C}]_{n,m}
    \end{aligned}
\end{equation}

In our proposed Algorithm. \ref{ap-sta-greedy}, $\mathcal{I}_{\mathbf{D}}$ denotes the index set of the matrix $\mathbf{D}$. (a) This greedy algorithm does not guarantee optimality, but only feasibility as a baseline for the problem formulated above, which has a time complexity of $\mathcal{O}(N^2M^2)$. (b) Thereafter, an optimal solution is proposed, we define an incidence matrix $\mathbf{A}_{g} \in \{0,1\}^{(M+N) \times |E|}$, where $|E|=N \times M$ denotes the total number of edges and $M+N$ denotes the total number of APs and STAs. The incidence matrix has the following structure according to our design:
\begin{equation}
\begin{aligned}
    \mathbf{A}_{g} = \begin{bmatrix}
                    \mathbf{A}_{M}\\
                    \mathbf{A}_{N}
                    \end{bmatrix},
\end{aligned}
\end{equation}
where
\begin{equation}
    \begin{aligned}
          \mathbf{A}_{M} = \begin{bmatrix}
                    1 \dots 1~0 \dots 0~\dots 0~\dots 0\\
                    0 \dots 0~1 \dots 1~\dots 0~\dots 0\\
                    \vdots\\
                    0 \dots 0~0 \dots 0~\dots~1 \dots1\\
                    \end{bmatrix} \in \{0,1\}^{M \times |E|},
    \end{aligned}
\end{equation}  
and 
\begin{equation}
    \begin{aligned}
          \mathbf{A}_{N} = \begin{bmatrix}
                    1~0\dots 0~1~0 \dots 0~\dots~ 1~0 \dots 0\\
                    0~1 \dots 0~0~1 \dots 0~\dots~0~1 \dots 0\\
                    \vdots \\
                    0~0\dots 1~0~0\dots 1~\dots~0~0\dots 1\\
                    \end{bmatrix} \in \{0,1\}^{N \times |E|}.
    \end{aligned}
\end{equation}  
Hence, one can conclude that $\mathbf{A}_{g}$ is {\em totally unimodular} (TU) that can be characterized as follows: 
\begin{itemize}
\item
Every entry in $\mathbf{A}_{g}$ is $0$, $+1$, or $-1$.
\item
Every column of $\mathbf{A}_{g}$ contains at most two non-zero entries ($+1$ or $-1$).
\item
if two non-zero entries in a column of $\mathbf{A}_{g}$ have the same sign, then the corresponding row of one is in $\mathbf{A}_{N}$, and the other in $\mathbf{A}_{M}$.
\item 
If two non-zero entries in a column of $\mathbf{A}_{g}$ have the opposite sign, then the corresponding rows are both in $\mathbf{A}_{N}$ or $\mathbf{A}_{M}$.
\end{itemize}
Fact: Every square, non-singular submatrix of a TU matrix is unimodular. Equivalently, every square submatrix of a TU matrix has determinant $0$, $+1$ or $-1$. 

One can vectorize $\mathbf{D}$, $\mathbf{X}$ in terms of $\mathbf{x} \in [0,1]^{|E|}$ and $\mathbf{d} \in \mathcal{R}^{|E|}$, and used to express the following equivalent optimization problem:
\begin{align}
        &\argmax_{\mathbf{x}} \mathbf{d}^{T}\mathbf{x}\\
        s.t.~~~ &~~~ \begin{bmatrix}\mathbf{A}_{g}\\ -\mathbf{A}_{g}\end{bmatrix}\mathbf{x} \leq \mathbf{b},~\mathbf{b} = \begin{bmatrix}
            \mathbf{1}_{M}\\ 
            \mathbf{R}\\
            \mathbf{0}_{M}\\
            \mathbf{R}
        \end{bmatrix}\\
            ~~~ &~~~ \mathbf{x} \in \{0,1\}^{|E|},
\end{align}
where $A_{g}$ represents the all possible connections between the AP group $\mathcal{N}$ and STA group $\mathcal{M}$ and $\mathbf{R} = [R(1),\dots,R(N)]^{T}$.
\begin{algorithm}[t]
\caption{Greedy Algorithm for AP-STA Pairing} 
\label{ap-sta-greedy}
$\mathcal{I}_{\mathbf{D}}$ = \text{sort elements of } $\mathbf{D}$ \text{ in decreasing order}\\
\textbf{For} descending element index  $\{n,m\}$ in $\mathcal{I}_{\mathbf{D}}$ \textbf{do}\\

~~~~$\mathbf{X}^{\prime} \leftarrow{} \mathbf{X}$, $\mathbf{X}^{\prime}_{n,m} \leftarrow 1$\;
~~~~\textbf{If}~$\mathbf{1}^{T}_{n}\mathbf{X}^{\prime}_{m} = 1$ and
    $\mathbf{1}^{T}_{m}(\tilde{\mathbf{X}_{n}^{\prime}})^{T} \leq R(n)$, $\forall m=1,\dots,M, n=1,\dots,N$~$\textbf{then}$
    
    ~~~~~~~~$\mathbf{X} \leftarrow{} \mathbf{X}^{\prime}$\;
    ~~~~\textbf{End If}\\
    \textbf{End For}
    
\end{algorithm}
\vspace{-3mm}
\begin{lemma}\label{integral}
Let $\mathbf{A}$ be a TU $m \times n$ matrix and let $\mathbf{b} \in \mathbb{Z}^{m}$. Then each vertex of the polyhedron
\begin{equation}
    P:=\{\mathbf{x} | \mathbf{A}\mathbf{x} \leq \mathbf{b}\}
\end{equation}
is an integer vector.
\begin{proof}
Refer to 8.1. in \cite{schrijver2012course}.
\end{proof}
\end{lemma}

It is well known that a linear programming (LP) problem with bounded constraints must have a vertex solution. Hence, according to Lemma \ref{integral}, \textbf{Problem \ref{Prob:AP-STA}} can be equivalently converted to LP because $A_{g}$ is a TU matrix and can force the vertices of the constraint in the form of a polyhedron to be an integer. According to the property of TU operation, $\mathbf{A} = \begin{bmatrix}
\mathbf{A}_{g}\\-\mathbf{A}_{g}\end{bmatrix}$ is also TU, as are $\begin{bmatrix} \mathbf{A} \\ \mathbf{I}\end{bmatrix}$ or $\begin{bmatrix} \mathbf{A} \\ -\mathbf{I}\end{bmatrix}$.  As the LP optimal solution is one of the vertexes of the constraint polyhedron, we define the following equivalent LP problem:
\begin{Prob}(Equivalent AP-STA Pairing: Linear Programming)\label{eq:AP-STA}
\begin{align}
        &\argmax_{\mathbf{x}} \mathbf{d}^{T}\mathbf{x} \nonumber \\
        s.t.~~~ &\mathbf{M}\mathbf{x} \leq \mathbf{b}, ~~\mathbf{M} = \begin{bmatrix}\mathbf{A}_{g}\\ -\mathbf{A}_{g} \\ \mathbf{I}_{|E|} \\ -\mathbf{I}_{|E|}\end{bmatrix},~ \text{and} ~~\mathbf{b} = \begin{bmatrix}
        \mathbf{1}_{M}^{T}\\
        \mathbf{R}\\
        \mathbf{0}_{M}^{T}\\
        \mathbf{R}\\
        \mathbf{1}^{T}_{|E|}\\
        \mathbf{0}^{T}_{|E|}
    \end{bmatrix} 
\end{align}
\end{Prob}
\begin{remark}
This Equivalent AP-STA pairing optimization can be solved optimally with LP algorithms that have the time complexity lower bounded by $\mathcal{O}((N\times M)^{2+\frac{1}{6}}\log(\frac{N \times M}{e}))$ \cite{cohen2019solving}, where $e$ denotes the relative accuracy. Since this problem can be solved and verified within polynomial time, problem 2 is an NP problem. Although the proposed greedy algorithm does not guarantee global optimality, it has lower time complexity. Therefore, the greedy algorithm can still be applied in practice to achieve faster implementation with compromised performance. Please note that the channel data rate matrix $\mathbf{C}$, available radio links for all APs and STAs $R(n)$ and $r(m)$ are assumed to be collected as the input data on APC.
\end{remark}

\subsection{Step 2: Radio Link Allocation}
After the AP-STA pairing is completed, we consider the subsequent radio link allocation problem. Suppose $\sum_{n=1}^{N}R(n) \geq M$, the first-stage AP-STA pairing should be established such that each STA can be covered by the AP. Hence, it is intuitive to use greedy algorithms to make full use of MLO, i.e., every STA's radio links $r(m)$ are exploited. This is because if the radio links are reused for connection between APs and STAs, where the reused number of a radio link is directly related to the collision rate and the maximum achievable data rate. This is unavoidable in contention-based Wi-Fi networks. The maximum number of radio links that all STA associates with its associated AP can be obtained from $\mathbf{\tilde{R}}_{n}=\sum_{m=1}^{M}\mathbf{X}_{n,m}\mathbf{r}(m)$. For example, $\mathbf{\tilde{R}}_{n}$ denotes the maximum number of radio links all STA associated with the AP $n$ can request. Denote the link selection matrix as $\mathbf{S} \in \{0,1\}^{F \times |E|}$
The radio link allocation problem can be formulated as follows: 
\begin{algorithm}[t]
\caption{Greedy Algorithm using PF scheduler} 
\label{greedy}
\textbf{For} each radio link edge $e$
\begin{equation}
\Phi^{cur}_{f} \xleftarrow{} \left( 1 - \frac{1}{T} \right)\Phi^{pre}_{f} + \frac{\mathbf{S}_{f}\mathbf{C}_{f}^{T}}{T(\mathbf{S}_{f}\mathbf{1}_{e}^{T})}\;
\end{equation}
where $T$ is the number of time slots we do average for the data rate, normally $T = 100$.\\ 
~~$\mathcal{F} = \{1,\dots,F\}$, $\mathbf{F} = [~]$\;
~~\\
~~\textbf{While} $\mathcal{F} \neq \{\}$\\
~~~~1. Append $f=\argmax_{\mathcal{F}}\{ P_{t}\left( f \right)\}$ to $\mathbf{F}$
where
\begin{equation}\label{eq:pf_allocation}
P_t(f) = \frac{\mathbf{S}_{f}\mathbf{C}_{f}^{T}}{\Phi^{cur}_{f}(\mathbf{S}_{f}\mathbf{1}_{e}^{T})}\;
\end{equation}
~~~~2. $\mathcal{F} \leftarrow{} \mathcal{F}\setminus{f}$\;
~~$\textbf{End While}$\\
~~\\

~~$\textbf{For}$ $i=1,\dots,F$\\
~~~~$f^{*} = \mathbf{F}[1]$\;
~~~~\textbf{If} $\sum_{f=1}^{F} \sum_{m=1}^{M} S_{f,n(M-1)+m} \leq \mathbf{\tilde{R}}_{n} ~ \text{for all} ~ n$ when $\mathbf{S}_{f^{*},e} = 1$  $\textbf{then}$ \\

    ~~~~~~~~~~~~$\mathbf{S}_{f^{*},e} \leftarrow 1,~\Phi^{pre}_{f} \leftarrow \Phi^{cur}_{f}$\;

~~~~\textbf{Else}\\
~~~~~~~~~~~~$\mathbf{F} \leftarrow \mathbf{F}[2:F]$ \;    
~~~~\textbf{End If}\\
~~\textbf{End For}
\end{algorithm}

\begin{Prob}(Radio Link Allocation: Integer Programming)
    \begin{align}
    &\argmax_{\mathbf{S}} \sum_{f=1}^{F} \log(\Phi^{cur}_{f})\\
    s.t.~~~&\textbf{card}(\mathbf{S}_{1:F,n(M-1)+1:nM}) \leq \mathbf{\tilde{R}}(n) ~ \text{for all n} \label{eq:card}\\
        ~~~&\Phi^{cur}_{f} = \frac{\mathbf{S}_{f}\mathbf{C}_{f}^{T}}{T(\mathbf{S}_{f}\mathbf{1}_{e}^{T})} + (1-\frac{1}{T})\Phi^{pre}_{f}\\
        ~~~&\mathbf{S} \in \{0,1\}^{F \times |E|}.
    \end{align}
\end{Prob}
The formulation aims to maximize the network throughput considering the proportional fairness of the channel condition and the previous throughput on each channel by introducing $\log(\bullet)$ into the objective functions, which is a common way of introducing proportional fairness to the variable of interest \cite{hao_tcom}. The current and previous average throughput per device of T slots on channel $f$ is denoted as $\Phi^{cur}_{f}$ and $\Phi^{pre}_{f}$ respectively. Note that the cardinal constraint in Eq. \eqref{eq:card} can also be characterized as a linear constraint, i.e., $\sum_{f=1}^{F} \sum_{m=1}^{M} S_{f,n(M-1)+m} \leq \mathbf{\tilde{R}}(n) ~ \text{for all} ~ n$.


Note that the integer programming problem formulated can be solved efficiently with our proposed greedy Algorithm \ref{greedy}, whose convergence is guaranteed by the proof in Appendix \ref{pf_proof}. The algorithm maximizes the aggregate utility for each arriving user inspired by the PF scheduler shown in Algorithm. \ref{greedy}. The complexity of this algorithm is $\mathcal{O}(FMN)$.

\begin{figure}[h]
    \centering
    \includegraphics[width = 0.43\textwidth,height=0.23\textwidth]{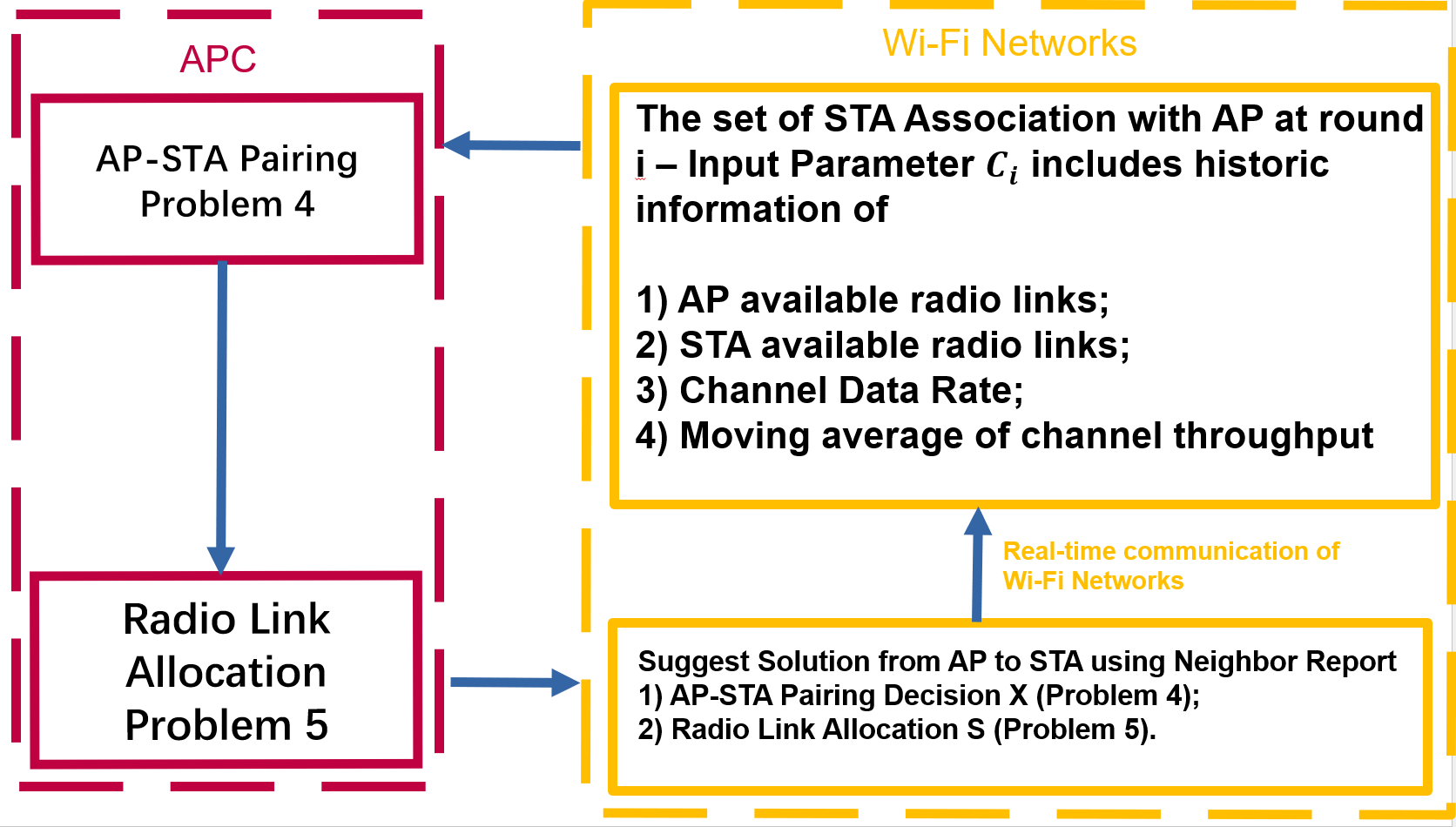}
    \caption{Flow Chart of the Proposed Method.}
    \vspace{-0.5cm}
    \label{fig:flowchart}
\end{figure}

\begin{remark}
The overall proposed method is shown in Fig. \ref{fig:flowchart} as flowchart and pseudo-code in the Algorithm \ref{overall_algorithm}: 1) APC collects STA association information as input from the Wi-Fi networks; 2) Solve the AP-STA pairing problem to obtain the optimal followed by the radio link allocation problem; 3) AP suggests the BSSID, that is, the MAC address of the specific radio link of interest, to STA using Neighbor Report following IEEE 802.11k\footnote{When STA scan the spectrum and finds out the BSSID with the highest RSSI is not in the suggested solution in neighbor report, STA does not follow the suggested solution but associate with the BSSID with the highest RSSI. This method can keep completing the channel rate matrix $\mathbf{C}$ to help the proposed algorithm reach a more accurate solution.}. It is noteworthy that the solution is delivered periodically since the channel condition varies, i.e., we can set periodicity to the coherence time of the channel, which is 0.978s in 11ax \cite{rappaport2002wireless}. Therefore, periodicity must be set on the scale of seconds, in which the overhead to deploy our proposed solution to Wi-Fi networks is negligible. Please note that it takes hours to complete one round of human-in-the-loop network management in the current enterprise-level APC/WLAN controller.
\end{remark}

\begin{algorithm}[t]
\caption{Pseudo-code of the Overall Algorithm} 
\label{overall_algorithm}
~~~~1. APC collect parameters from Wi-Fi network: $R(n)$, $r(m)$, $\mathbb{C}_{f,n,m}~\forall f,m,n$, $\Phi^{pre}_{f}~\forall f$; \\\

~~~~2. Solve Problem 4 with $R(n)$, $r(m)$, and $\mathbb{C}^{t}_{f,n,m}$ as input using LP algorithm (Interior Point Method) ; \\\

~~~~3. The solution $\mathbf{X}$ to Problem 4 and the historic throughput $\Phi^{pre}_{f}$ are used as the input to Algorithm 2; \\\

~~~~4. Recommend $\mathbf{X}$ from Step 2 and $f^{*}$ for each radio link edge $e$ indicated by $\mathbf{X}$ from Step. 3 to Wi-Fi networks using Neighbor Report; \\\

~~~~5. Repeats 1-4 with a period of T; \
\end{algorithm}

\section{Performance Evaluation}
\label{simulation}
In this section, we evaluate the performance of the proposed AP-STA pairing and radio link allocation algorithm in terms of throughput and proportional fairness metrics under various PHY-MAC features such as SNR, MCS, channel efficiency, the number of MLD coexisting in the same channel, etc. The simulation is based on the implementation of the PHY-MAC cross-layer channel data rate. To this end, the PHY abstraction method is used to simulate the packet error rate (PER) for IEEE channel model B\cite{11axChannel}. To this end, we first introduce the PHY abstraction for the packet error rate of IEEE channel model B. Then, we implement the improvised Bianchi model for the IEEE 802.11 DCF basic \cite{haowifi} and the PHY rate corresponding to various PHY-MAC parameters. Then, based on the simulated scenarios and simulated cross-layer channel data rate, we apply our proposed algorithms and examine the acquired data rate versus SNR w.r.t. multiple MCS and convergence analysis.

\subsection{Data Rate Estimation over Wi-Fi Channels}

Estimation of data rate for Wi-Fi channel comprises two-folds: 1) Packet Error Rate (PER) estimation aided by PHY abstraction; 2) MCS data rate estimation corresponding to PHY-MAC features. The combination of estimation of the PER and the PHY-MAC data rate yields the actual channel data rate of each radio link that any AP-STA pair can provide. 

\subsubsection{Packet Error Rate Estimation}

PHY layer abstraction that generates a packet error model can be utilized for fast and accurate PER estimation, which is an efficient deployment for the proposed AP-STA pairing and the radio link allocation algorithm. The PHY layer abstraction avoids the hour-long runtimes that the traditional full PHY simulation suffers from. 

To realize PHY layer abstraction, we apply the Exponential Effective SNR Mapping (EESM), which provides an effective signal-to-noise ratio (ESNR) \footnote{ESNR \cite{ieeephy1,ieeephy2,yin2024adr} translates the SNRs of all sub-carriers in the frequency-selective channel into a single scalar value. ESNR is used to estimate the instantaneous PER by looking up from a pre-stored PER-SNR table for the IEEE channel model B \cite{9149444}.} of each packet over the entire channel for any radio link \cite{sian2021ns3, sian2022TCOM}:
\begin{equation}
\tilde{\gamma}=-\beta \ln \left(\frac{1}{N_{sc}} \frac{1}{N_{ss}} \sum_{i=1}^{N_{sc}} \sum_{j=1}^{N_{s s}} \exp \left(-\frac{\gamma_{i, j}}{\beta}\right)\right),
\end{equation}
where $N_{sc}$ is the number of sub-carriers, $N_{ss}$ is the number of the spatial streams, and $\beta$ is the tuning parameter for optimization \cite{sian2022TCOM}. $\gamma_{i, j}$ is the SINR at the $i$th sub-carrier for the $j$th stream, which is given by
\begin{equation}
    \gamma_{i, j} = \frac{S_{i,j}}{I^{s}_{i,j}+N_{i,j}},
\end{equation}
where $S_{i,j}$ is the received signal power, $I^{s}_{i,j}$ is the inter-stream interference, and $N_{i,j}$ is the noise power. Denote the obtained PER as $P_{e} = f_{\text{table}}(\tilde{\gamma})$, where $f_{\text{table}}(\bullet)$ denotes the PER-SNR lookup table used in the following section for actual channel data rate calculation. 

\begin{figure}[ht]
\centering
 \subfigure[SISO case]
{\includegraphics[width=0.4\textwidth]{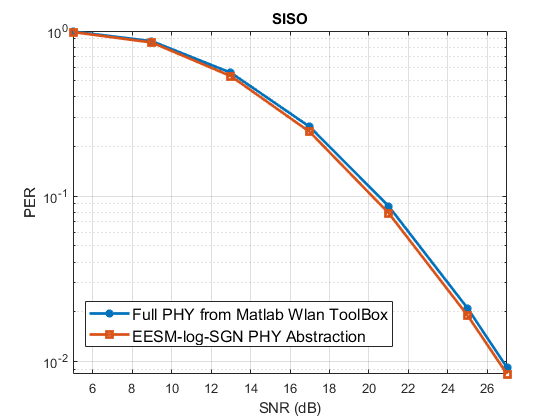}}
\label{fig:SISO}
\centering
\subfigure[MIMO case]{
\includegraphics[width=0.4\textwidth]{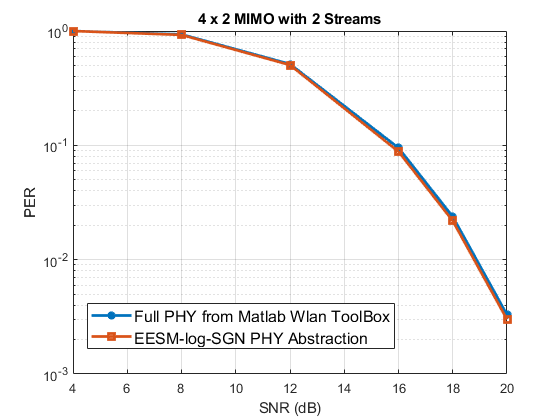}}
\label{fig:MIMO}
\caption{PHY Abstraction Validation: IEEE Channel Model-B, MCS 3, LDPC, Channel Bandwidth = 20 MHz, Packet Size = 1500 Bytes.}
\label{fig:PHYValidation}
\end{figure}

The physical layer feature PER vs SNR based on various MCS setting affecting overall throughput is achieved using EESM-log-SGN PHY abstraction method in our previous publication \cite{wcl, wns3_liu}. This has been validated to align well with the simulated PHY-layer PER and data rate based on the Matlab WLAN ToolBox with very high accuracy up to above 99\%. Our EESM-log-SGN PHY abstraction method relies on the offline simulation data based on Matlab WLAN ToolBox and then characterizes those data by only using a few parameters (called log-SGN parameters) which are further stored in our python simulator for fast simulations with high PHY accuracy. As Fig. \ref{fig:PHYValidation} shows, to validate the PER vs SNR based on our PHY abstraction method, we tested both SISO and MIMO scenarios under a specific setup. As a result, the simulated PER vs SNR curve in both cases well fit with the reliably simulated curves based on WLAN ToolBox.

\begin{figure}[h]
    \centering
    \includegraphics[width = 0.42\textwidth,height=0.25\textwidth]{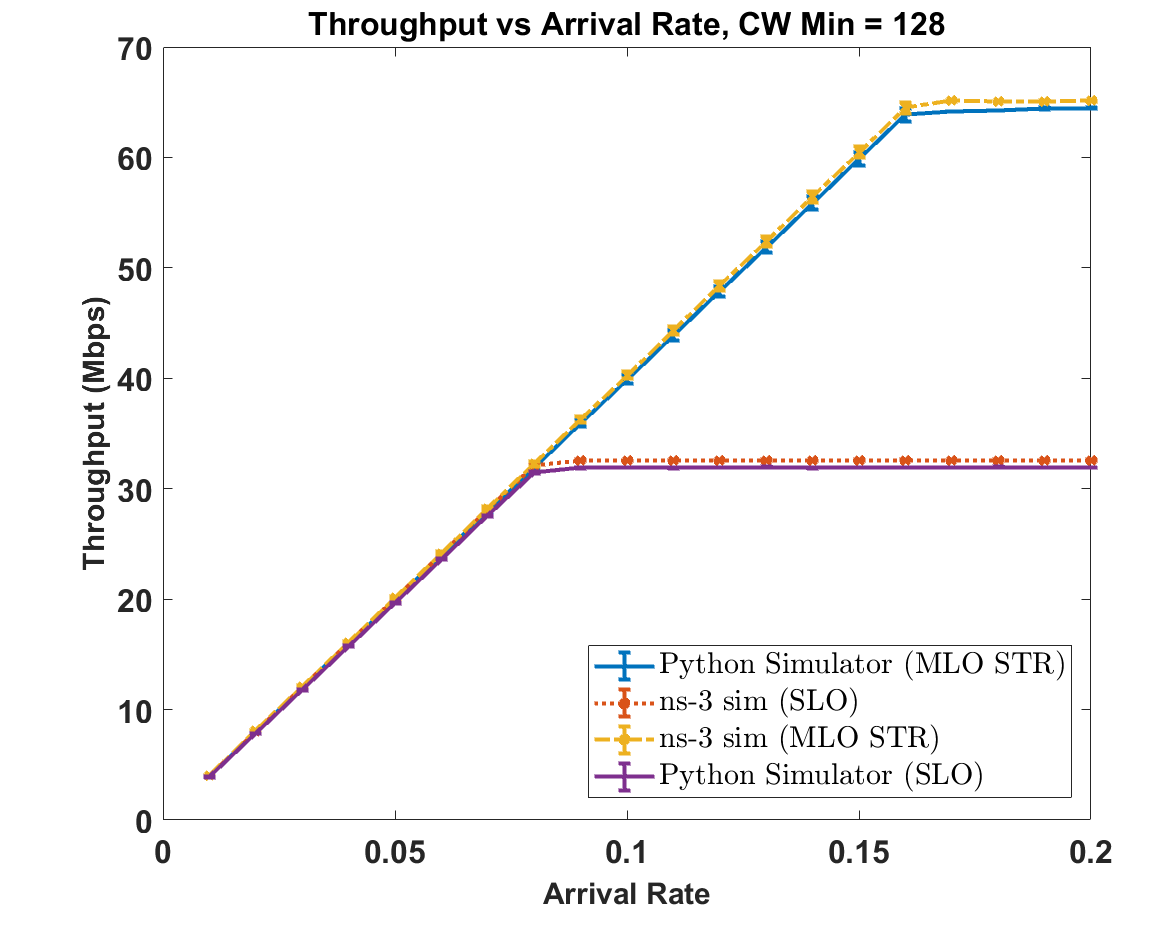}
    \caption{Throughput vs Arrival Rate: Initial contention window size is 128, the number of AP is 1 and station is 20 and each device has 2 radio links in MLO scenario and 1 radio link in SLO scenario, MCS 6 without aggregation. Bandwidth is 20 MHz for SLO scenario and $20 + 20$ MHz for MLO scenario. SNR = 20dB.}
    \label{fig:throughput_arrival_rate}
    \vspace{-0.5cm}
\end{figure}

\subsubsection{Channel Data Rate}
In this subsection, we focus on the analytical model of the throughput efficiency of the IEEE 802.11 DCF \cite{bianchi2000performance, haowifi} as the channel access method and the MCS data rate for any radio link on the $2.4$ GHz, $5$ GHz, and $6$ GHz bands to calculate the channel data rate. The channel data rate can be expressed as the multiplication of the normalized throughput and MCS data rate:
\begin{equation}
    \begin{aligned}
        C = P_{e} \times Tpt \times D^{mcs},
    \end{aligned}
\end{equation}
where $C$ represents the element of the channel data rate matrix $\mathbb{C}_{f,n,m}$ w.r.t different MCS, PER corresponding to various SNR, different choices of bandwidth, varying coding rate, etc. The analytical model of normalized throughput $Tpt$ and the MCS data rate $D^{mcs}$ is shown in Appendix \ref{analytical}.

\begin{figure*}[t] 
  \centering
  
  \subfigure[MCS=3] {\includegraphics[width=.27\textwidth,height=0.2\textwidth]{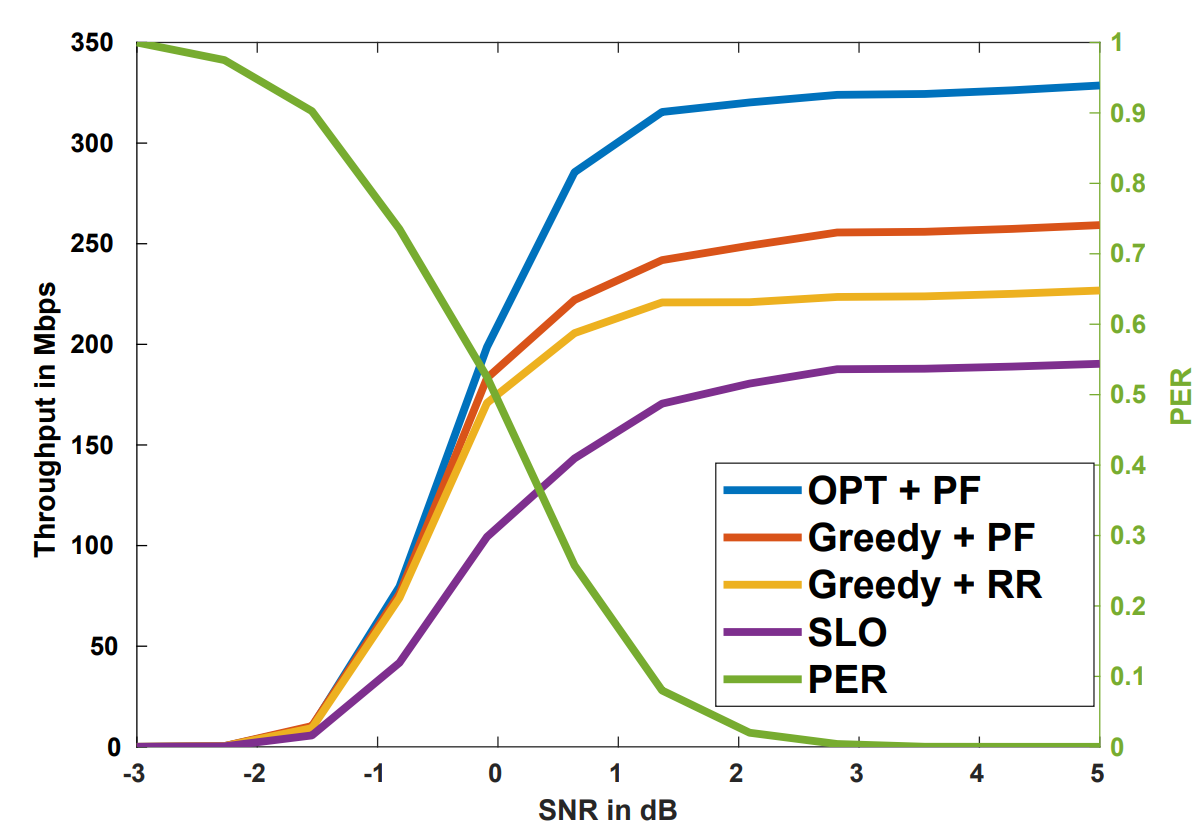}\label{fig:performance1}}
  \quad
  \subfigure[MCS=6] {\includegraphics[width=.27\textwidth,height=0.2\textwidth]{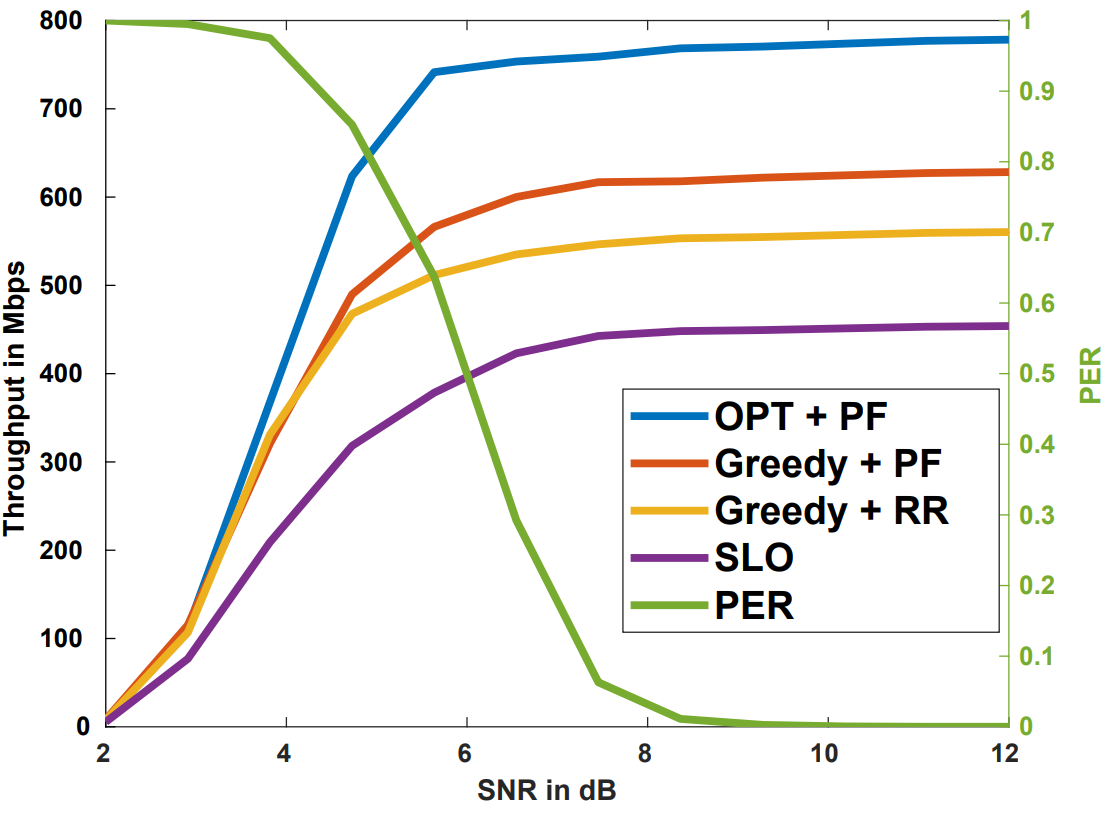}\label{fig:performance2}}
  \quad
    \subfigure[MCS=9] {\includegraphics[width=.27\textwidth,height=0.2\textwidth]{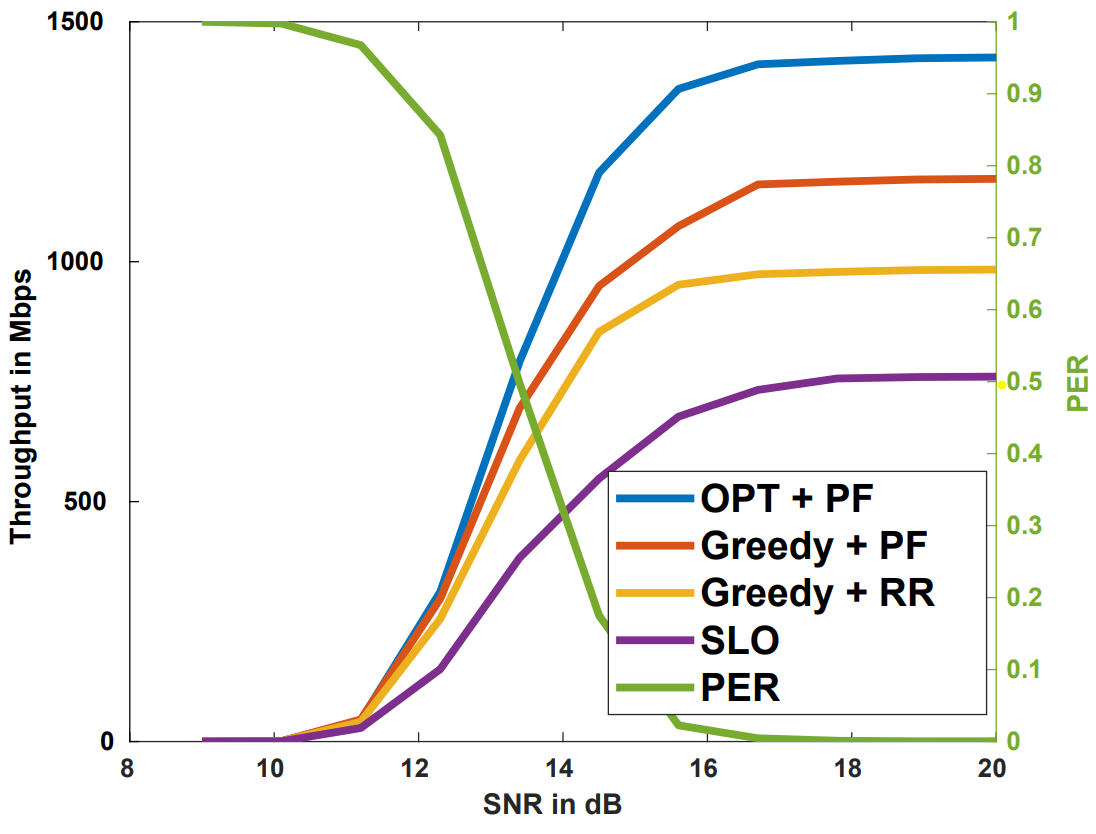}\label{fig:performance3}}\quad
  
  \caption{PHY-MAC Network Throughput vs SNR for three algorithms: 1) proposed optimal AP-STA pairing algorithm plus PF algorithm; 2) proposed greedy algorithm plus PF algorithm; 3) proposed greedy algorithm plus RR algorithm. PER is shown on the right side to show the corresponding varying SNR.}
  \vspace{-0.5cm}
  \label{fig:performance}
  
\end{figure*}

\subsection{Performance Analysis: Throughput}
The performance evaluation is implemented on a computer with a single CPU (Intel(R) Core(TM) i7-9700K) and implemented using Python and cvxpy toolkit as the LP solver. Our python simulator is validated by the latest ns3 SLO and MLO features in Fig. \ref{fig:throughput_arrival_rate}. In Fig. \ref{fig:throughput_arrival_rate}, MLO with 2 radio links has 2x arrival rate compared with SLO and reaches 2x throughput than SLO when saturated. Python simulator MLO STR mode is shown to align with ns3 simulator MLO STR mode with high accuracy. Please note that as arrival rate increases, network will be saturated and throughput plateau at the peak. Arrival rate is the standard Poisson arrival rate that represents the probability of packet arrival event in each slot time in WiFi system. In previous art \cite{FrmaVTC2020}, python simulator with SLO operation is also validated by Bianchi Model in terms of network throughput. The PHY rate is validated against IEEE channel model B in Matlab WLAN Toolbox. We extended the design to cross-layer throughput to support the MLO and AP controller features.

\begin{figure}[h]
    \centering
    \includegraphics[width = 0.42\textwidth,height=0.25\textwidth]{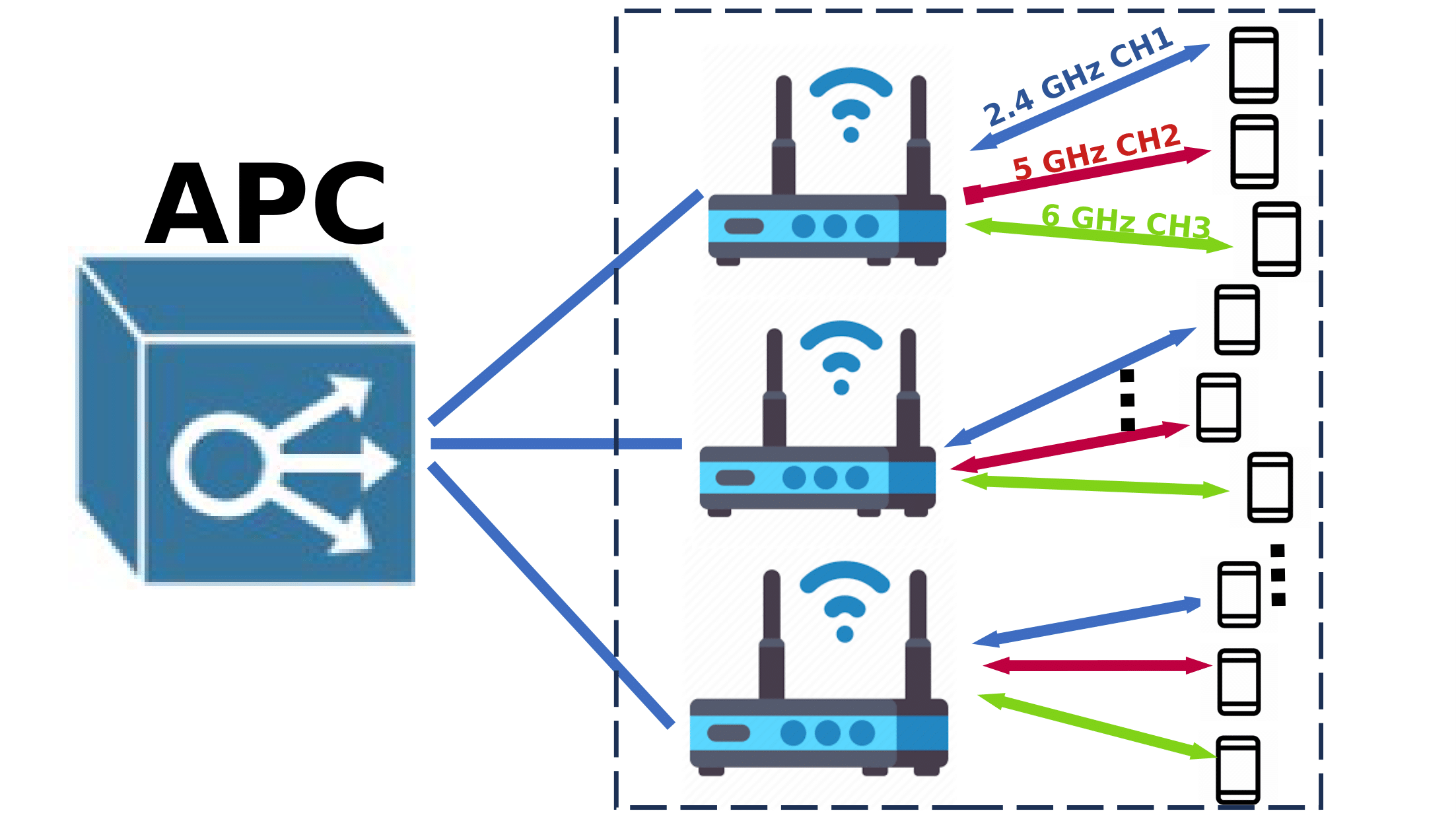}
    \caption{Network Scenario.}
    \label{fig:sim_setup}
    \vspace{-0.5cm}
\end{figure}

As shown in Table. \ref{tb:parameter}, the network scenario comprises $40$ MHz bands on $2.4$ GHz, $80$ MHz bands on $5$ GHz, and $160$ MHz bands on $6$ GHz for independent data transmission, which is chosen as the maximum bandwidth typically allowed in each band \footnote{On the 5 GHz channel, we do not consider the dynamic frequency selection (DFS) channel; thus the maximum available bandwidth is 80 MHz.}. We instantiate $3$ MLD APs and $15$ MLD STA and in a total of $3$ overlapping basic service set (BSS). This network scenario is shown in Fig. \ref{fig:sim_setup}. The modulation and coding scheme (MCS) varies from $\{3,6,9\}$ \footnote{The MCS is set to a constant value to assess the throughput of various algorithms. This fixed setting of MCS acts as a control variable, ensuring that throughput converges to a certain value and enabling a consistent, horizontal comparison across different algorithms.}. Each sub-carrier within a channel has the same modulation and coding rate. For MCS $\{3, 6, 9\}$, we have the corresponding data rate as $\{34.4, 68.8, 144.1\}$, $\{77.4,154.9,324.3\}$, and $\{114.7,229.4,480.4\}$ on $40$, $80$, and $160$ MHz respectively according to Eq. \eqref{eq:datarate} with parameters in \cite{ieeephy3}. The performance results are summarized from $100$ rounds of Monte Carlo trials. 




\begin{table}[t]
\caption{Parameters for Wi-Fi networks}
\label{tb:parameter}
\centering
\resizebox{.42\textwidth}{!}{%
\begin{tabular}{|c|c|}
\hline
\textbf{Parameters} &\textbf{Value} \\                                  \hline
Carrier frequency           & 2.4 GHz, 5 GHz, 6 GHz \\ \hline
Bandwidth & $40$ MHz, $80$ MHz, $160$ MHz   \\ \hline
Channel Type & IEEE channel model B \\ \hline
slot time ($\mu$s)                        & $9$    \\     \hline
SIFS ($\mu$s)                        & 16                                   \\ \hline
DIFS ($\mu$s)                        & 34                                   \\ \hline
EIFS ($\mu$s)                        & SIFS + ACK + DIFS                                   \\ \hline

PHY preamble \& Header ($\mu$s)     & 20                                   \\ \hline
Payload Size                   &  1500 Bytes                                \\ \hline
ACK\_Timeout ($\mu$s)                 & 300 
 \\ \hline
ACK (Bytes)                          & 14 + PHY Header                      \\ \hline
$CW_{min}$                          &  $16$                      \\ \hline
$CW_{max}$                           & $1024$                    \\ \hline
$m$                          & $6$                    \\ \hline

\end{tabular}%
}
\end{table}

In Fig. \ref{fig:performance}, we evaluate performance for four algorithm, i.e., OPT $+$ PF, Greedy $+$ RF, Greedy $+$ RR, and SLO. Out of four evaluated algorithms, OPT $+$ PF is the proposed algorithm that delivers the highest throughput performance. Greedy $+$ PF represents the combination of the greedy algorithm proposed in Algorithm \ref{ap-sta-greedy}, which is used to showcase the potential performance gain one can obtain from the optimal solution.  Regarding the greedy $+$ RR algorithm, the traditional Round-Robin (RR) algorithm replaces the proposed PF algorithm to demonstrate the performance gain degradation. SLO represents the network performance when all devices have a single link with a proportionally fair AP-STA pairing decision, which is used to show the advantages of MLO. The optimal and the greedy algorithms have similar performance when the SNR is low, and the optimal algorithm plus PF outperforms the greedy algorithm plus PF and greedy algorithm plus Round-Robin (RR) significantly when the SNR is large by up to 27.84\% and 45.07\%. RR is a traditional resource allocation algorithm used in modern communication networks \cite{hao_tcom}. RR algorithm allocates each radio link $e$ with the channel number sequentially, e.g., radio link 1 is allocated in channel 1 and radio link 2 is allocated in channel 2, etc. Please note that the performance gap between the optimal and the greedy algorithm emerges relatively early in the higher MCS value. This phenomenon can be observed by comparing Fig. \ref{fig:performance1} and \ref{fig:performance3}. This is because higher MCS introduces more susceptibility to the channel model and consequently more variations to the achievable data rate. In other words, the resource allocation becomes more complicated as MCS grows. Therefore, the optimal solution still remains the best performance; however, it is challenging for the greedy solution to explore feasible solutions for a more complicated problem with higher MCS. The SLO is evaluated as the baseline to the algorithms proposed for MLO. In the SLO case, all devices are single-link devices. Each AP is paired with $5$ STAs and AP $1$, $2$, and $3$ operates on $40$, $80$, and $160$ MHz respectively. The performance of the SLO is the worst because, unlike the SLO, MLO allows AP to communicate different STAs on different radio links simultaneously in which each STA can communicate with the optimal channel data rate.       

In summary, we can conclude that the proposed algorithm is able to achieve significant performance gain for 11be network throughput with multiple MLDs.

\subsection{Performance analysis: Fairness}
\begin{figure*}[t] 
  \subfigure[PF Algorithm] {\includegraphics[width=.23\textwidth,height=.19\textwidth]{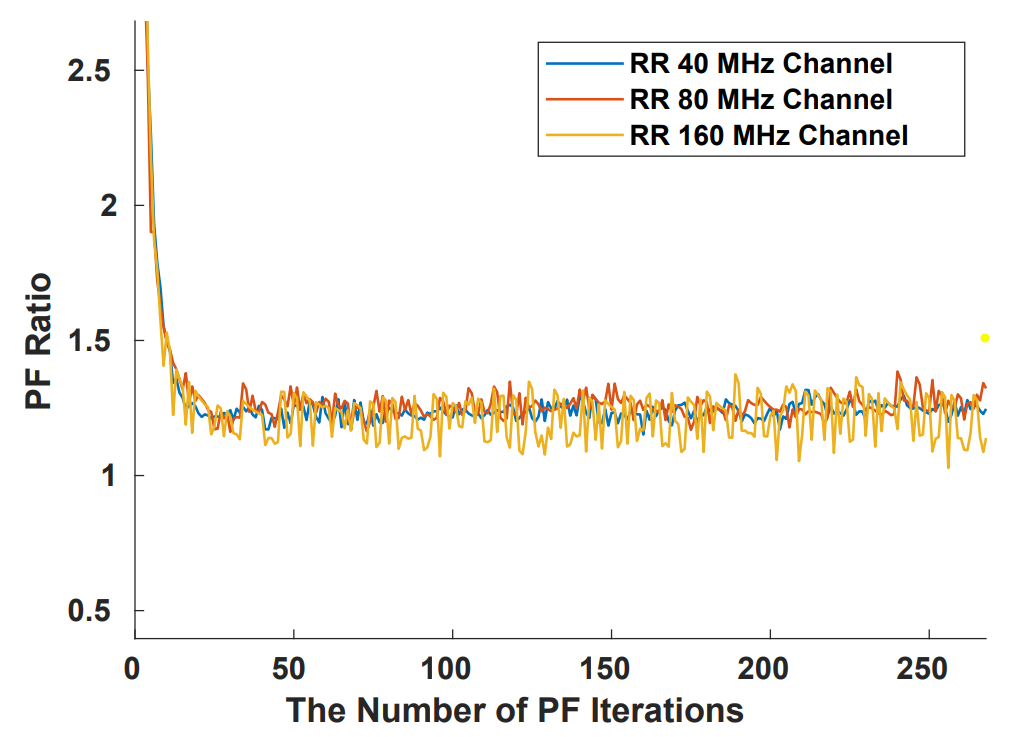}\label{fig:pfratio1}}
  \quad
  \subfigure[RR Algorithm] {\includegraphics[width=.23\textwidth,height=.19\textwidth]{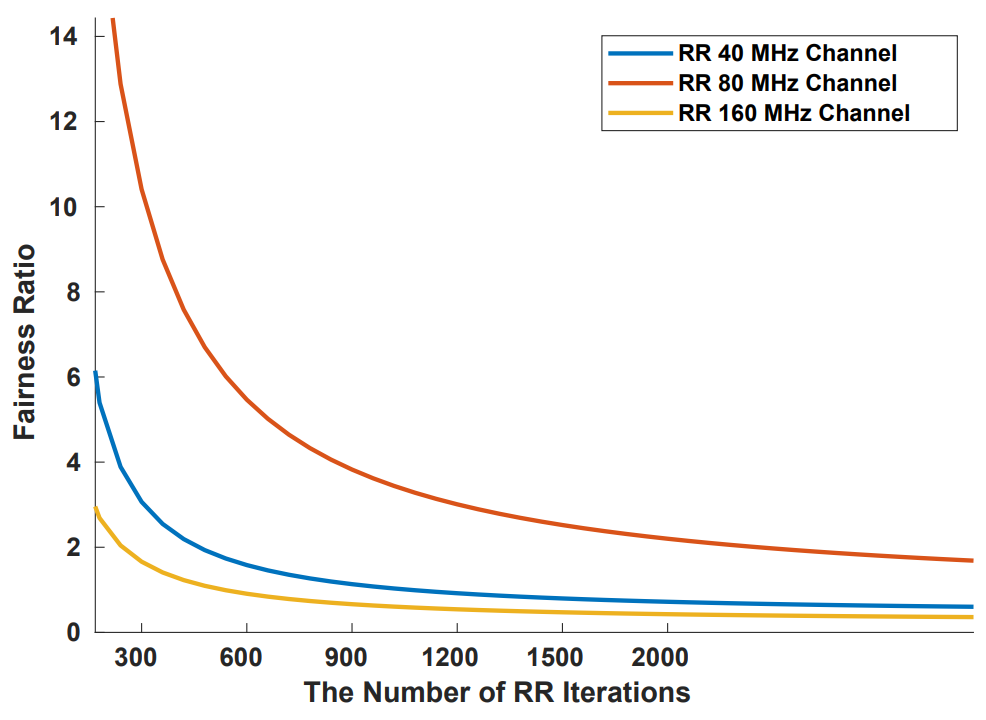}\label{fig:rrratio1}}
  \quad
    \subfigure[PF Algorithm] {\includegraphics[width=.23\textwidth,height=.19\textwidth]{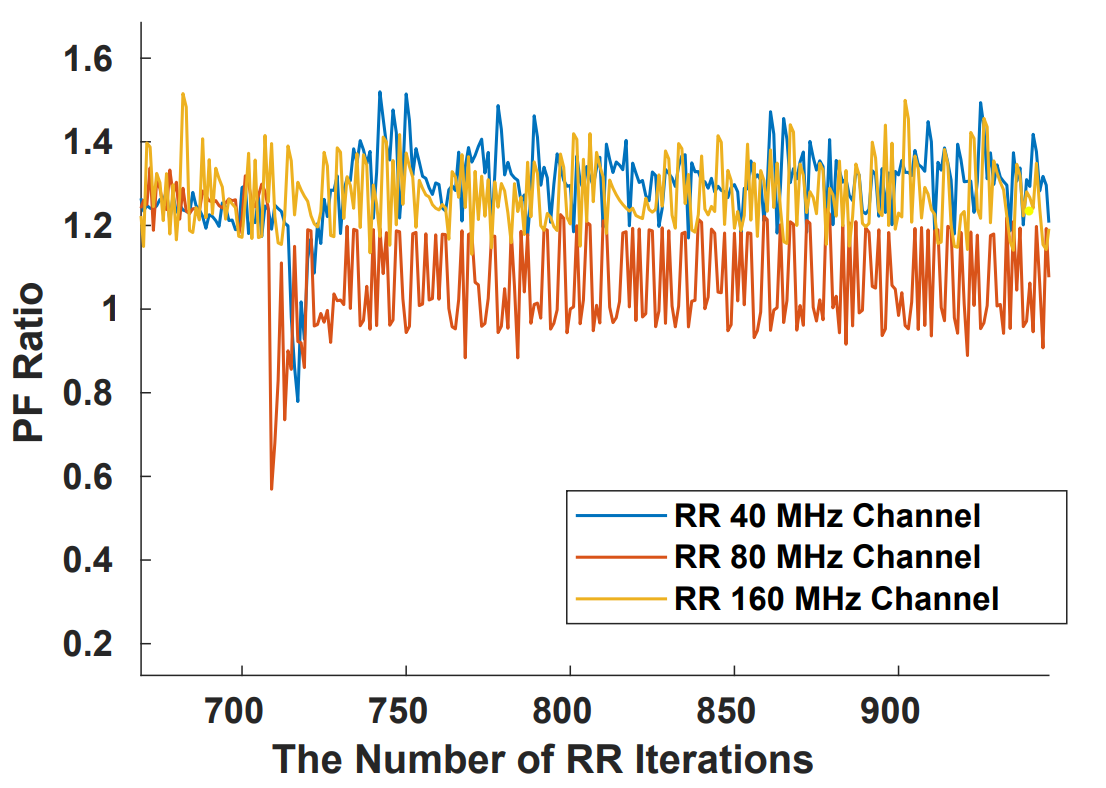}\label{fig:pfratio2}}\quad
 \subfigure[RR Algorithm] {\includegraphics[width=.23\textwidth,height=.19\textwidth]{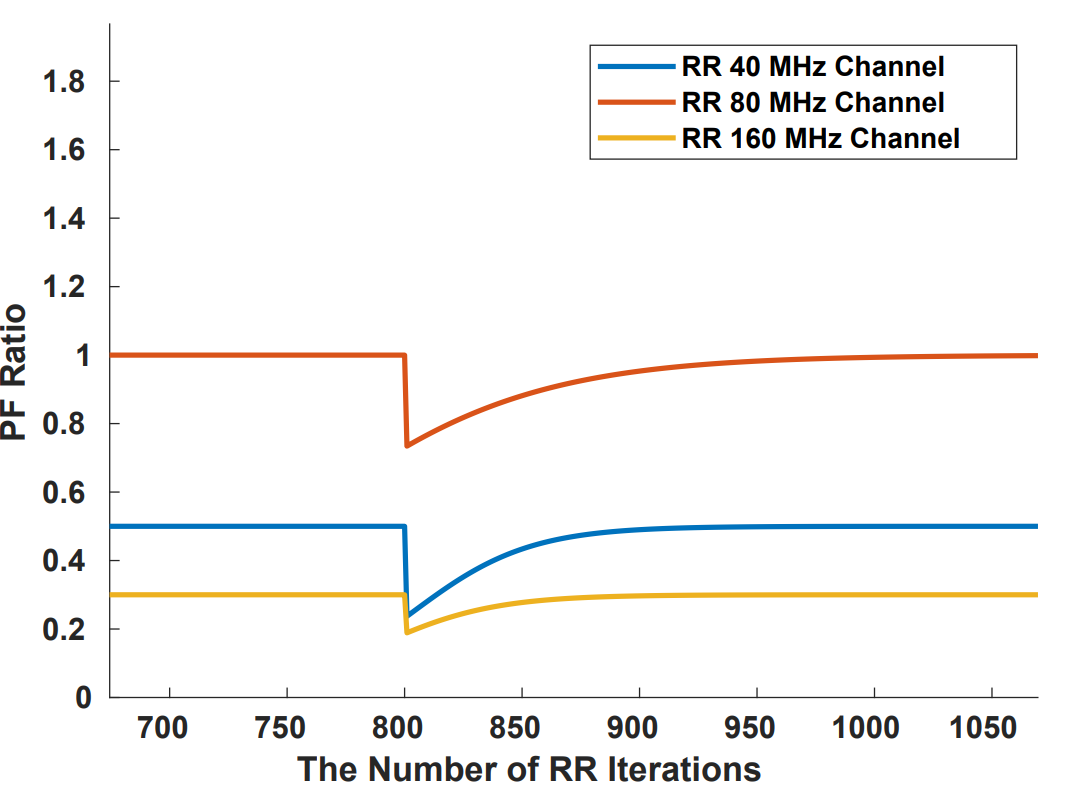}\label{fig:rrratio2}} \quad
  \caption{Fairness Ratio vs The Number of Iterations. MCS is $9$ in (a) and (b) and MCS is changed from $9$ to $6$ in (c) and (d).}
  \label{fig:rate PF}
  \vspace{-0.5cm}
\end{figure*}

The proposed PF algorithm is designed to strike a balance between maximizing total throughput and ensuring equitable resource distribution, especially in scenarios where users experience varying channel conditions. This mechanism is critical in dense deployments, where interference is a significant challenge, as it can lead to unfair channel allocation to radio links, disproportionately affecting users with weaker signals. Our algorithm aims to mitigate this unfairness by dynamically adjusting resource allocation, ensuring that users in adverse conditions are not unduly penalized. In this subsection, we provide additional mathematical and empirical evidence to elucidate how our approach not only optimizes overall network performance but also maintains fairness among all users, regardless of their individual channel conditions and the varying levels of interference they may encounter.

Next, we demonstrate the performance gain from the radio link allocation algorithm in detail. The performance evaluation is designed firstly by choosing the MCS for all APs to be $9$ and then the MCS is dropped to $6$ to compare the performance of the Round-Robin (RR) and the PF scheduler in Algorithm \ref{greedy} in terms of data rate and proportional ratio. The RR allocation algorithm is also called the max-min fairness method, which allocates the radio link edge based on the channel bandwidth; for example, the sequential allocation of $4$, $2$, $1$ slice portion of the radio link edges to the $160$, $80$, $40$ MHz channel, respectively.

\begin{figure}[t] 
  \centering
  \subfigure[PF Rate] {\includegraphics[width=.35\textwidth,height=.22\textwidth]{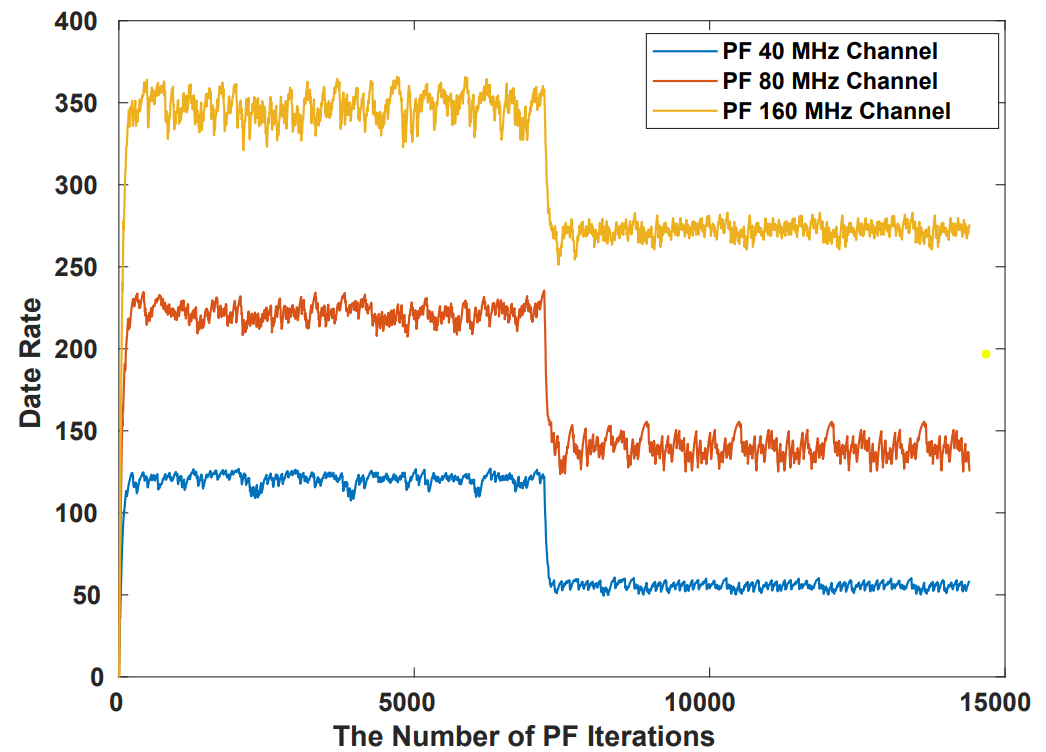}\label{fig:pfrate}}
  
  \subfigure[RR Rate] {\includegraphics[width=.35\textwidth,height=.22\textwidth]{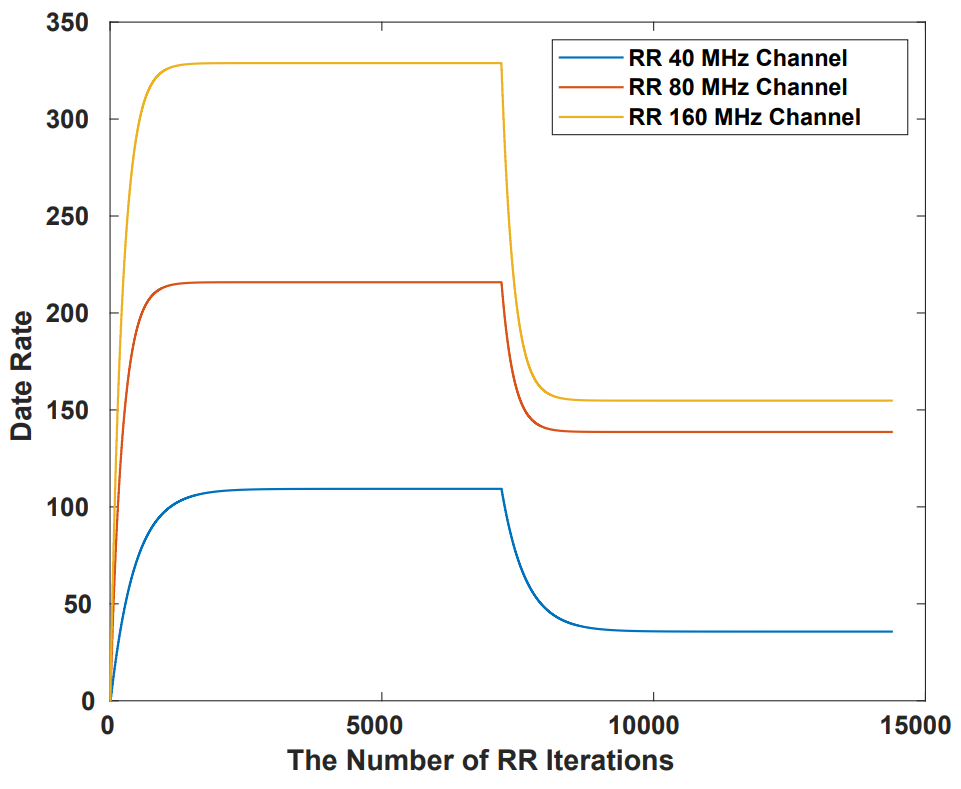}\label{fig:rrrate}}
  \caption{Data Rate vs The Number of Iterations: Comparison between PF and Max-min fairness.}
  \label{fig:rate}
\end{figure}

In Fig. \ref{fig:rate}, The data rate vs the number of iterations results based on two different radio link allocation approaches are compared. In Fig. \ref{fig:pfrate}, one can observe that PF can converge within less than 15 iterations while, in Fig. \ref{fig:rrrate}, RR can only converge within almost 2000 iterations. Moreover, the PF radio link allocation can maintain a higher data rate than RR by the average of $11.4$\%, $9.8$\%, and $18.5$\% on radio links $40$ MHz, $80$ MHz, and $160$ MHz respectively when MCS is $9$ and $71.8$\%, $3.2$\%, and $30.7$\% on radio links $40$ MHz, $80$ MHz, and $160$ MHz respectively when MCS is changed to $6$. Although it seems that RR has a smoother curve than PF, it is noteworthy that RR cannot maintain a satisfying performance on the $160$ MHz radio link when a sudden change from MCS $9$ to $6$ happens, which also takes much longer response time to converge compared to PF. This performance evaluation indicates that the PF has higher stability and faster convergence than RR, and most importantly, it can reach a higher network throughput.

In summary, suppose that, in a densely overlapping Wi-Fi network, STA adopts passive scanning with fast initial link setup (FILS) from AP, which acts as a mini-beacon transmitted every 20 ms. Considering the additional airtime for the probing, authentication, and association process. The whole connection establishment process can take at least 40 ms. In such a case, it will take hundreds of milliseconds for the PF algorithm to converge because each PF iteration requires one reassociation for STA. This process will be much faster and more practical than RR.

The performance evaluation presented above has showcased the data rate performance of the two methods; however, the fairness among radio links remains to be explored. We first express the fairness ratio as follows:
\begin{equation}
    P_t(f) = \frac{\mathbf{S}_{f}\mathbf{C}_{f}^{T}}{\Phi^{cur}_{f}(\mathbf{S}_{f}\mathbf{1}_{e}^{T})},
\end{equation}
where $P_t(f)$ represents the ratio of the instantaneous data rate and the average data rate on the radio link $f$. If the instantaneous data rate aligns with the average data rate, then one can conclude that the radio link enters stability. This metric describes the fairness shared by each radio link, and the closer the value of fairness ratios among radio links, the more fairness the algorithm can guarantee the network. This also explains why the maximum value of $P_t(f)$ out of all $f = 1, \dots, F$ is chosen as the radio link $e$ at time slot $t$, as shown in Eq. \eqref{eq:pf_allocation}. In Fig. \ref{fig:pfratio1}, it is shown that the fairness ratio of PF method converges within 10 iterations and each radio link has a similar fairness ratio. On the other hand, in Fig. \ref{fig:rrratio1}, the convergence rate is very slow while each channel shares quite different fairness ratios based on the RR method. Fig. \ref{fig:pfratio1} and \ref{fig:rrratio1} showcase the fairness ratio of two methods that are based on the optimal AP-STA pairing solution under MCS $9$. To further demonstrate the dynamics of the proposed algorithms, the MCS is changed to $6$ and the response of two methods is shown in Fig. \ref{fig:pfratio2} and Fig. \ref{fig:rrratio2}. As shown in Fig.\ref{fig:pfratio2}, the fairness ratios return to stability in 20 rounds. The proportional fairness w.r.t. the change of MCS is quickly restored. However, in Fig.\ref{fig:rrratio2}, the fairness ratios go through a sudden drop and return to the original value, which is still considered worse than the proposed solution in terms of proportional fairness. This is because the RR method sequentially allocates radio link resources and can only utilize the lower MAC layer information such as the number of radio links without the PHY layer information such as channel data rate to maintain fairness from the cross-layer perspective.

\begin{figure}[h] 
  \centering
  \subfigure[Performance Evaluation] {\includegraphics[width=.22\textwidth]{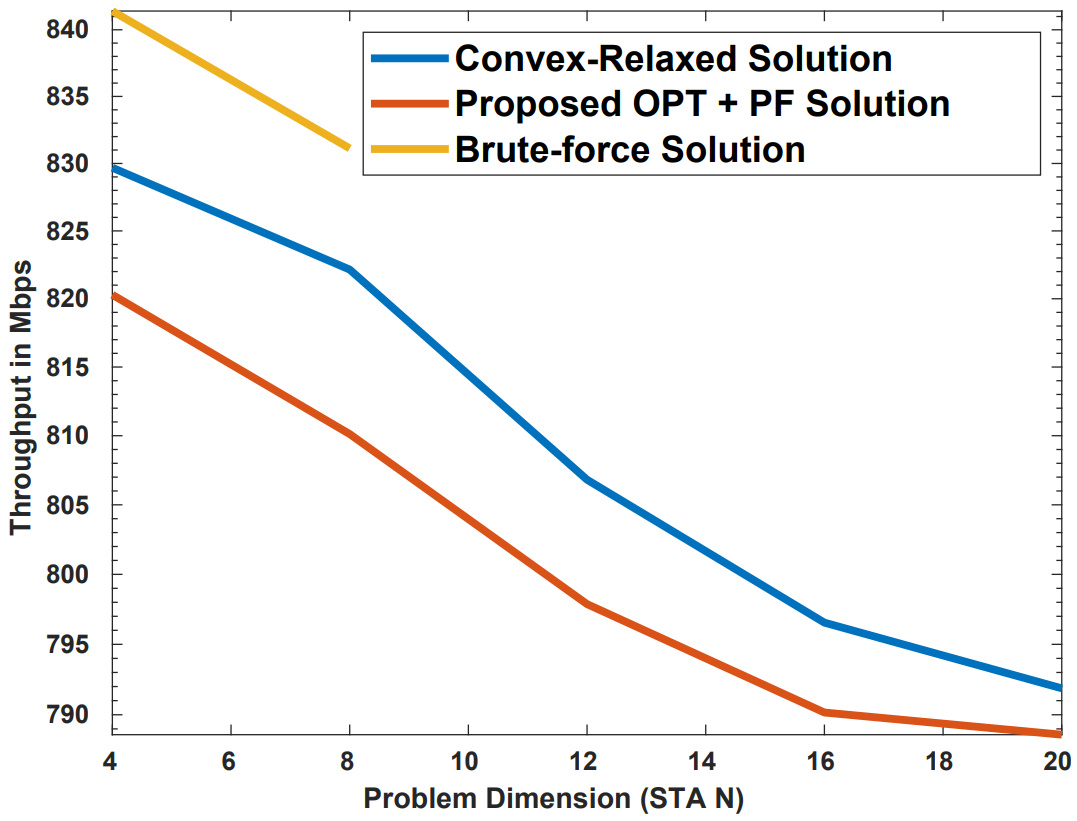}\label{fig:joint2}}
  \subfigure[Running Time] {\includegraphics[width=.22\textwidth]{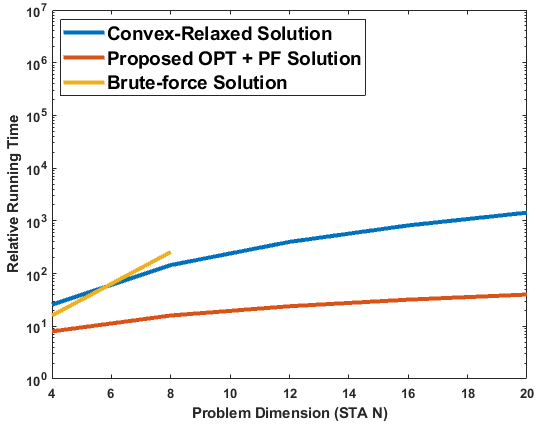}\label{fig:joint1}}
  \caption{Performance and Time Complexity Comparison for the Joint Problem Solution and the Proposed Solution.}
  \label{fig:joint}
\end{figure}
\subsection{Comparison between the Solution to Joint Problem 1 and Our Proposed Solution}
In this subsection, we set another network scenario (2 APs) for comparison between the solutions to Problem 1 and our proposed solution. In this scenario, we still set the $40$, $80$, and $160$ MHz operating channels on the $2.4$, $5$, and $6$ GHz, which are the same setup in Table. \ref{tb:parameter}. MCS $6$ is used, and SNR is randomly chosen from $[6,9]$ dB.

The convex-relaxed solution solves Problem 1 by relaxing it to a penalized convex problem with only the sub-optimal solution without guarantee of the exact solution \cite{zhang_tcom}. The brute-force solution to Problem 1 can deliver the global optimum. However, as shown in Fig. \ref{fig:joint2}, the time complexity of the Brute-force solution grows exponentially as the number of STA increases, and the convex-relaxed solution also consumes much more running time than our proposed algorithm by almost two degrees of magnitude. In Fig. \ref{fig:joint1}, the Throughput vs Problem dimension is shown. One can observe that the brute-force can only deliver the solution when the number of STA is less than or equal to $8$ on our computer. Since the brute-force algorithm loops through all possible combinations to reach the global optimum, the time complexity is exponential, as introduced in Problem 1. In other words, when the problem dimension grows, it takes an impractical run-time for the brute-force algorithm to yield the solution. Both the Convex-Relaxed Solution and Brute-force perform better than our proposed solution with only marginal gains, which both consume much more running time.

\subsection{Summary}
In this section, the performance analysis of our proposed algorithm is mainly divided into two subsections: throughput and fairness. Since our proposed algorithm comprises two parts including AP-STA pairing and radio link selection, either part of our proposed algorithm is replaced with a baseline algorithm to show its irreplaceable superiority in terms of the network throughput. In the fairness performance analysis, we use a fairness ratio as the metric to quantify the fairness reached by the proposed algorithm and the baseline algorithm, in which the proposed algorithm is shown to have much faster convergence with satisfying fairness in the crowded network.

\section{Conclusion and Future Work Discussion}\label{sec:conclusion}

In this paper, we propose for the 11be network with MLO and APC a novel AP-STA pairing algorithm and a radio link allocation method to achieve network throughput maximization with guaranteed proportional fairness. The performance of the proposed algorithms is then evaluated via an analytical channel data rate model considering the PHY-MAC features. Performance evaluation shows that the AP-STA pairing plus the PF scheduler can outperform the baselines in terms of network throughput, fairness ratio, convergence, and time complexity. In future work, we aim to investigate the joint throughput and latency optimization in the Wi-Fi 7 network emphasizing the MLO and the potential application of deep reinforcement learning.





\appendix

\subsection{Proof of Asymptotic Proportional Fairness}\label{pf_proof}
We show that the greedy algorithm considering PF maximizes the aggregate throughput while guaranteeing the asymptotic PF on one channel. Denote $p^{t}_{f,e}$ as the probability of the radio link $e$ at time slot $t$ being assigned with channel $f$. Assume the channel data rate does not change within $t$ slots, then we have the average data rate
\begin{equation}
    \bar{D}(f) =  \sum_{e} \sum_t p^{t}_{f,e}\mathbf{C}^{t}_{f,e}.
\end{equation}

Therefore, the network utility maximization problem \cite{srikant2013communication} is
\begin{align}
&\max ~ \sum_{f}{\log(\sum_{e} \sum_t p^{t}_{f,e}\mathbf{C}^{t}_{f,e})}   \label{eq:utility} \\
&\text{s.t.} ~\sum_{e}\sum_{f} p^{t}_{f,e} \leq 1,~
p^{t}_{f,e} \geq 0, ~\forall t, e.
\end{align}
Note that the above problem is a convex problem since the log function with the composition of an affine function still preserves concavity. Applying Lagrange multipliers, we obtain the following:
\begin{equation}
    \begin{aligned}
    \sum_{f}{\log( \sum_{t} \sum_{e} p^{t}_{f,e}\mathbf{C}^{t}_{f,e})}- \sum_{t} \lambda_{t}(\sum_{e}\sum_{f} p^{t}_{f,e}-1)
    \end{aligned}
\end{equation}

Taking the derivative w.r.t. $p^{t}_{f,e}$, we obtain the optimal solution as 

\begin{equation}
    \frac{\mathbf{C}^{t}_{f,e}}{\bar{D}(f)} - \lambda^{*} = 0 \quad \text{if} \quad p^{t}_{f,e} > 0,
\end{equation}
Asymptotically, the PF algorithm helps the AP network to reach the PF, i.e., 
\begin{equation}\label{eq:PF_alg}
    \lim_{t \xrightarrow{} \infty}\frac{\bar{D}(1) }{\mathbf{C}^{t}_{1,e}} = \dots = \lim_{t \xrightarrow{} \infty}\frac{\bar{D}(F) }{\mathbf{C}^{t}_{F,e}},~ \forall e.
\end{equation}
Hence, the allocation method shown in Eq. \eqref{eq:pf_allocation} follows the optimal condition of the 11be network utility maximization problem with PF. 

\subsection{Analytical Model of Normalized Throughput and MCS Data Rate}\label{analytical}
 Denote the probability of at least one MLD transmitting as $P_{tr}$, the successful transmission, i.e., no collisions, as $P_{s}$. With the PER $P_{e}$ obtained from the previous section, we can obtain the normalized throughput $Tpt$ \cite{haowifi} for each channel : 
\begin{equation}\label{eq:s}
 \begin{aligned}
 &Tpt= \\ &\frac{(1-P_{e})P_{s}P_{tr}E[Pkt]}{(1-P_{tr})\sigma + P_{tr}P_{s}(1-P_{e})T_{s} + P_{tr}(1-P_{s})T_{c} + P_{tr}P_{s}P_{e}T_{e}}
 \end{aligned}
\end{equation}
where $\sigma$ is the slot duration and $T_{s}$, $T_{c}$, $T_{e}$ represent the duration for a successful transmission, collision, and PHY error due to the bad channel condition, respectively. 
The probability that at least one node
transmits in a slot is denoted as $P_{tr}$ and the successful
transmission probability is denoted as $P_{s}$. Due to page limit, please refer to \cite{haowifi} for expressions of $P_{tr}$ and $P_{s}$. Please note that the PHY error duration is the same as the collision duration because both result in a packet error. The transmission duration and the collision duration are calculated in the following Eq. \eqref{eq:ts} and \eqref{eq:tstc}:
\begin{align}
T_{s}&= H + E[Pkt] + \text{SIFS} + \delta + \text{ACK} + \text{DIFS} + \delta, \label{eq:ts}\\
T_{c}&= H + E[Pkt] + \delta + \text{EIFS},  
\label{eq:tstc}
\end{align}
where $E[Pkt]$ is the mean frame duration of the packets. Here $H$ represents the MAC and PHY header time, and $\delta$ is the propagation delay equal to 0.1 $\mu s$.   Short Inter-frame Space ($\text{SIFS}$), $\text{DIFS}$ and Extended Inter-frame space ($\text{EIFS}$) are time duration specified in 802.11 standards required for a wireless interface to process a received frame and respond with a response frame. The MCS data rate $D^{mcs}$ is shown as follows:
\begin{equation}
\label{eq:datarate}
D^{mcs}=\frac{N_{S D, U} * N_{B P S C S, U} * r_{c}}{T_{D F T}+T_{GI}},
\end{equation}
where $N_{S D, U}$ denotes the number of data subcarriers per resource unit. $N_{B P S C S, U}$ is the number of coded bits per subcarrier per stream for the resource unit, and $r_c$ is the coding rate. The values
of those parameters are defined in IEEE standard \cite{802.11ac} for
different spatial streams and MCS. $T_{D F T}$ and $T_{GI}$ are the OFDM symbol duration ($3.2$ $\mu$s) and guard interval duration ($0.8$ $\mu$s).

\bibliographystyle{IEEEtran}
\bibliography{reference}
\end{document}